\documentclass[preprint]{aastex}

\shorttitle{Non-equilibrium ionization code and its application} 
\shortauthors{Ji, Wang and Kwan}

\begin{document}
\newcommand{\ssst}{\scriptscriptstyle}

\title{NON-EQUILIBRIUM IONIZATION MODEL FOR STELLAR CLUSTER WINDS AND ITS
APPLICATION}

\author{Li Ji,  Q. Daniel Wang, and John Kwan}
    
\affil{Department of Astronomy, University of Massachusetts, Amherst,
MA 01003-9305; ji@nova.astro.umass.edu, wqd@astro.umass.edu,
and kwan@nova.astro.umass.edu}

\begin{abstract}
   We have developed a self-consistent physical model for super 
stellar cluster winds based on combining a 1-D steady-state 
adiabatic wind solution and a non-equilibrium ionization calculation.
Comparing with the case of collisional ionization equilibrium, we find that
the non-equilibrium ionization effect is significant in the regime of a high
ratio of energy to mass input rate and manifests in a stronger soft X-ray
flux in the inner region of the star cluster. 
Implementing the model in X-ray data analysis softwares (e.g., XSPEC) 
directly facilitates comparisons with X-ray observations. Physical
quantities such as the mass and energy input rates of stellar 
winds can be estimated by fitting observed X-ray spectra. The fitted 
parameters may then be compared with independent measurements from 
other wavelengths. Applying our model to 
the star cluster NGC 3603, we find that the wind accounts 
for no more than 50\%  of the total ``diffuse" emission,  and the derived mass
input rate 
and terminal velocity are comparable to other empirical estimates.
The remaining emission most likely originate from numerous low-mass
pre-main-sequence stellar objects.
\end{abstract}


\section{INTRODUCTION}
Super star clusters (SSCs) consist of densely-packed massive stars 
with typical ages of a few Myr. They are the most energetic coeval 
stellar systems, identified in a wide range of star-forming galaxies 
\citep{whit00}. Within a characteristic radius of a few pc or less,  winds from 
individual stars in such a cluster  are expected to collide and merge into a 
so-called stellar cluster wind. In some extragalactic systems, such 
as the nuclear region of M82 \citep{stri02}, winds from 
star clusters, either individually or collectively, can be energetic
enough to drive galaxy-scale superwinds. Consequently, cluster winds 
can play an important role in shaping star formation and galaxy 
evolution in general \citep{heck90}.

Observationally,  X-ray emission that appears to be diffuse has been detected 
around star clusters (e.g., NGC 3603, \citealt{moff02};
Arches cluster at the Galactic center, \citealt{yusef02}, \citealt{rock05},
\citealt{wang06}). Theoretically, 
both stellar cluster wind \citep[e.g.,][and references therein]{stha03}  
and galactic superwind \citep[e.g.,][]{ stst99, brsc99} have been investigated,  
particularly in terms of their expected X-ray luminosities. 
However, neither  the observed spectral nor spatial information 
has been used to confront 
the theoretical model prediction quantitatively. Furthermore, most models assume 
collisional ionization equilibrium (CIE) and no non-equilibrium ionization (NEI) 
model is available for a quantitative comparison with current 
X-ray data of stellar cluster winds and/or superwinds.

In this paper, we first introduce our non-equilibrium ionization model 
for stellar cluster winds based on combining a 
1-D steady-state adiabatic wind solution (\S \ref{sec-model}) with a 
non-equilibrium ionization calculation (\S \ref{sec-neqi}).  The non-equilibrium
ionization effect is examined within a grid of parameter space, which 
is described in \S \ref{sec-predictions}. Then we report its application to 
the diffuse X-ray emission of NGC 3603 (\S \ref{sec-application}). 
The physical parameters (e.g., mass input rate, terminal velocity, and  
metallicity)  of the stellar cluster wind are derived from the fits  and
compared with other independent measurements and empirical estimates.

\section{X-RAY EMISSION FROM STELLAR CLUSTER WINDS}
\label{sec-model}
On the scale of a galaxy,  supernovae in a starburst 
region can drive a strong wind. \citet{cc85}
describe it by an outflow model with uniformly distributed mass and energy
depositions within a radius.
Their solutions are scalable to the case of a star cluster, in which
the mass loading and energy input are due to stellar winds instead of supernovae.

Essentially analogous to  the model of \citet{cc85}, \citet{crr00} present an
analytic 1-D model to describe the cluster wind. In addition, their 
3-D numerical simulation of stellar winds from 30 stars produces a 
mean flow that  agrees well with their analytic model.
They predict that the X-ray emission from the cluster wind of   
the Arches cluster \citep[see also][]{raga01} is detectable \citep{lawyz04}.
Also based on the work of \citet{cc85}, \citet[][henceforth, SH03]{stha03}
present a model that allows for a lower energy transfer efficiency and mass 
loading. This model predicts global properties such as the X-ray luminosity and  
gas temperature at an individual cluster center. 
A crude comparison of the model predictions and existing {\it Chandra}
observations does not give any conclusive result as to whether or not the 
diffuse emission is genuinely associated with cluster winds. 
 
\begin{figure}[ht]
  \centerline{  
\includegraphics[height=0.7\textwidth,angle=90]{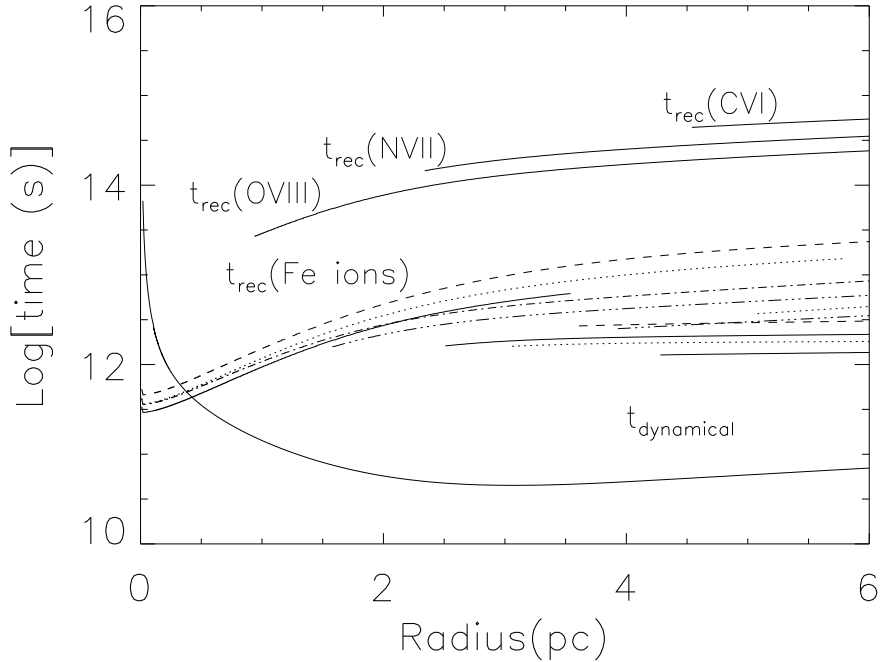}
    \vspace{-0.15in}
  }
  \caption{
    ~Dynamical timescale compared with intrinsic recombination
    timescales for a 1-D cluster wind with an exponentially distributed mass 
    input rate $\dot{M}_{0}=2.3 \times 10^{-4}\rm ~M_{\odot} ~yr^{-1}$, 
    a terminal velocity $V_{\infty}=2000 \rm ~km~s^{-1}$, and a sonic radius $r_{s}=1.38 \rm~pc$.}
  \label{fig-dyn_t}
\end{figure}

All of the above theoretical models  assume 
CIE for the ionization structure. However, one may suspect that the gas in a
cluster wind may be in a highly NEI state, because of both the rapid shock 
heating in the wind-wind collision and the fast adiabatic cooling in the 
subsequent expansion. Fig. \ref{fig-dyn_t} illustrates 
such a NEI condition in a cluster wind.  Clearly, the dynamic
(adiabatic) cooling time scale defined by $T/(dT/dt)$ is much shorter than the
recombination time scales of all key ion species,  and  CIE does not hold. 

% time scale comparison
%figure1 

Moreover, the mass and energy injections in the previous theoretical models are
input parameters and are crudely estimated from the properties (e.g., spectral type
and radio flux) of individual massive cluster stars.
Such a modeling approach suffers significant uncertainties and has little predictive power. 
Stellar mass loss remains
one of the largest uncertainties in stellar evolution \citep[e.g.,][]{kupu00}. 
Massive stars are also not the only sources of mass loss;   
other cold mass loading, such as outflows from protostars(e.g., SH03), may also be important. 

In the following subsections, we introduce our dynamic model of a stellar
cluster wind.

\subsection{Cluster Wind Dynamics}
\label{sec-dyn}
We consider a young star cluster (age $\leq 5\times 10^{6} ~\rm yrs$), 
from which the main mass and energy injections are due to stellar winds, 
especially from those massive OB and Wolf-Rayet stars.
Extending SH03's work, which  assumes a uniform distribution 
of stars, we have constructed a 1-D steady-state wind model with  (1) either 
an exponential or uniform distribution of stars; (2) a steady and spherically
symmetric cluster wind; 
(3) mass and energy injections following the stellar mass
distribution; (4) gravity and radiative cooling being neglected.
This cluster wind can then be described by the following steady-state Euler
equations:
      \begin{eqnarray}
      \frac{1}{r^{2}}\frac{d}{dr}(\rho ur^{2})=\dot{m}(r)
      \end{eqnarray}
      \begin{eqnarray}
      \rho u\frac{du}{dr}=-\frac{dP}{dr}-\dot{m}(r)u
      \end{eqnarray}
      \begin{eqnarray}
      \frac{1}{r^{2}}\frac{d}{dr}\left [\rho
ur^{2}(\frac{1}{2}u^{2}+\frac{\gamma}{\gamma -1}\frac{P}{\rho})\right ]=\dot{E}(r)
      \end{eqnarray}
where $u,\rho$ and $P$ are the velocity, density and pressure of the flow
respectively,  and  $\gamma$ is the adiabatic index, taken as $5/3$ for an ideal
gas, while $\dot{m}(r)$ and $\dot{E}(r)$ are the mass and energy input rates per
unit volume and are assumed to be proportional to the stellar mass density
$\rho_{s}(r)$. 
 
The total mass and energy input rates are  $\int_{0}^{\infty}4\pi
r^{2}\dot{m}dr=\dot{M_{0}}$ and $\int_{0}^{\infty}4\pi
r^{2}\dot{E}dr=\dot{E_{0}}$. 
When the stellar mass distribution is assumed to be uniform, $\dot{M_{0}}=\frac{4}{3}\pi
r_{c}^{3}\dot{m}$ and $\dot{E_{0}}=\frac{4}{3}\pi r_{c}^{3}\dot{E}$, where
$r_{c}$ is the cluster radius, which coincides with the sonic radius; 
the analytic solutions of Eq. (1)-(3) can be found in \citet{cc85}. 
For an exponential stellar mass distribution, $\rho_{s}(r)\propto exp(-r/r_{sc})$, 
where $r_{sc}$ is the scale radius. 
We solve the above Euler equations numerically. 
Because gravity is negligible, the sonic radius depends on the stellar
mass distribution alone. 

The above hydrodynamical equations do not predict how the electron 
temperature changes in the cluster.  The temperature depends on how the 
electron is heated.  
We consider two extreme cases: one that has  
the electron and ion temperatures in equilibration, i.e., $T_{e}=T_{i}=T$; 
the other that sets the
initial $T_{e}/T = 10^{-3}$ and lets the ratio evolve via electron-ion Coulomb collisions. 
Here, the mean temperature $T\equiv (\mu m_{p}/k)P/\rho$ and 
$\mu$ is the mean mass per particle in terms of the proton mass $m_{p}$. 
Assuming cosmic abundances, the variation in the electron temperature
follows the equation \citep[e.g., Eq. (1) in ][]{bork94}: 
    \begin{eqnarray}
     \frac{d (T_{e}/T)}{dt}=1.2\times 10^{-4} n (\frac{T}{10^{7}
K})^{-3/2}(\frac{T_{e}}{T})^{-3/2}(1-\frac{T_{e}}{T}) \rm ~yr^{-1},
    \end{eqnarray} 
where  $n$ is the total particle number density. 

\begin{figure}[ht]
      \centerline{
      \includegraphics[width=0.7\textwidth,angle=90]{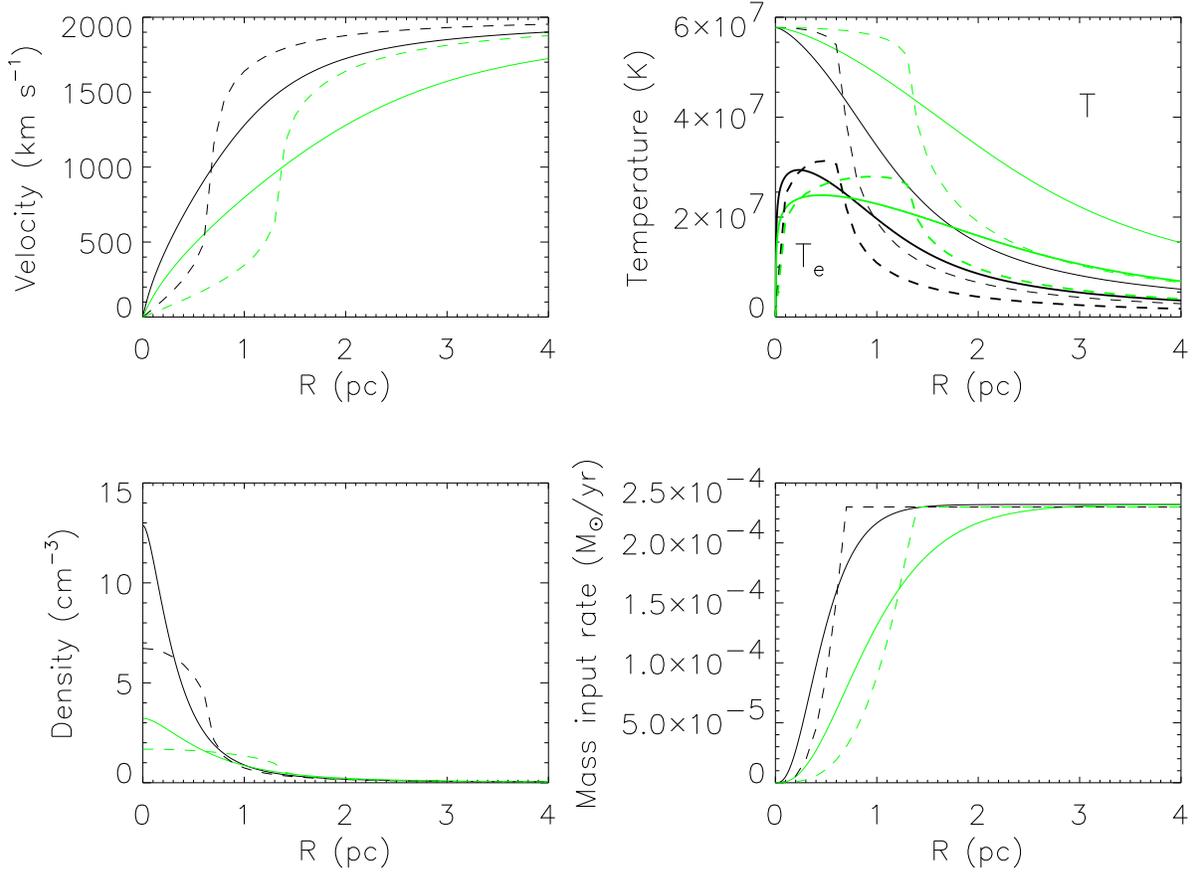}
       }
      \caption{
      The radial profiles of the cluster wind (solid line --- exponential
        distribution; dash line --- uniform distribution)
        the velocity (top left panel), mean temperature ($T$, top right,
       thin lines), electron temperature ($T_{e}$, top right, thick lines), density
      (bottom left), and cumulative mass input rate (bottom right). Two models 
        with different sonic radii ($0.69 \rm ~pc$, black; $1.38 \rm ~pc$, green)
       are shown for each stellar mass distribution. The same mass input rate
        $\dot{M}_{0}=2.3 \times 10^{-4}\rm ~M_{\odot} ~yr^{-1}$ and 
        stellar wind terminal velocity
       $V_{\infty}=2000 \rm ~km~s^{-1}$ are assumed. 
      }
      \label{fig-dyn}
    \end{figure}

Our model has four parameters: mass input rate $\dot{M}_0$, stellar wind terminal
velocity $V_{\infty}$ (assuming $\dot{E}_{0}=1/2\dot{M}_{0}V_{\infty}^{2}$),
abundance Z, and the scale radius $r_{sc}$ (or cluster radius
$r_{c}$  for a uniformly mass distributed model).  Fig. \ref{fig-dyn} 
illustrates the radial profiles of the mean/electron temperature, density,
terminal velocity and mass input rate for these two stellar distribution 
models with the same sonic radius.  Two sonic radii (0.69 pc, black; 
1.38 pc, green) are considered here.
When the sonic radius is larger, the density is lower in the inner part of 
the cluster, the temperature drops more slowly, and the terminal velocity of the
wind is reached at a larger radius.

\clearpage
%wind profiles: uniform and exponential
%figure2

\subsection{X-ray Emission Calculation}
\label{sec-obs}
The broad band X-ray luminosity $L_{X}$ of the cluster wind within a region
($r<R$) is given by:
   \begin{eqnarray}
     L_{X}=\int_{0}^{R} 4\pi r^{2}n_{e}n_{h}\int_{X}\Lambda_{\nu}d\nu ~dr
     \label{equa-lumi}
   \end{eqnarray}
where $X$ is the frequency band of interest; $n_{e}$ and $n_{h}$ are the
electron and 
hydrogen number density respectively; $\Lambda_{\nu}$ is the specific emissivity 
at frequency $\nu$,  and the cluster wind is assumed to be optically thin. Generally $\Lambda_{\nu}$ is a
function of elemental abundances, local ionic fractions and electron temperature. 

The model-predicted spectrum $S_{\nu}$ from a given projected annulus between the
inner-to-outer radii $[R_{in}, R_{out}]$ can be calculated using the following
equation:
    \begin{eqnarray}
    S_{\nu}(R_{in},R_{out})=\frac{1}{D^{2}}\int_{R_{in}}^{R_{out}} 2\pi R dR
\int_{R}^{\infty} \frac{2n_{e}(r)n_{h}(r)\Lambda_{\nu}r}{\sqrt{r^{2}-R^{2}}}dr, 
    \end{eqnarray}
where $D$ is the distance to the cluster. The observed surface brightness can be
predicted as:
    \begin{eqnarray}
   \Sigma_{X}(R)=\frac{1}{4\pi}\int_{R}^{\infty}\frac{2n_{e}(r)n_{h}(r)r}{\sqrt{r^{2}-R^{2}}}\int_{X}\int_{X\arcmin}\Lambda_{\nu\arcmin}A(\nu\arcmin)E(\nu\arcmin,\nu)d\nu\arcmin
d\nu dr
    \end{eqnarray}
where  $A(\nu\arcmin)$ and $E(\nu\arcmin,\nu)$ are the on-axis effective area and
 energy response matrix of the X-ray telescope/instrument. 

In principle, both the spatial and spectral information of the data can be used
to  constrain our
model. But while we may reasonably assume the spectrum of the undetectable 
point sources to be the same as that of detected point sources, a similar assumption 
about the surface brightness cannot be made because the
detected point sources are sparse and incomplete in a typical observation.
Therefore, unlike the spectrum, the observed surface brightness cannot easily be
compared with the model prediction at present. 

\section{NON-EQUILIBRIUM IONIZATION SPECTRAL MODEL}
\label{sec-neqi}
In the past, NEI models have been developed for (1)  time-dependent
isobaric or isochronic cooling of shock-heated gas \citep[e.g.,][]{sudo93}, (2)  
delayed ionization in SNRs \citep[e.g.,][]{bork01}, and (3)  delayed
recombination in galactic winds 
\citep[e.g.,][]{brsc99} or in the Local Bubble \citep[e.g.,][]{brei96}.  The 
atomic data used are 
typically more than  a decade old and are stored in a way that is difficult to update.
Moreover, the physical processes included are not complete,  due to either the
lack of atomic data  or 
the use of the CIE approximation.  One example of such a process is the
cascade of electrons following recombinations
into highly excited levels, which could strongly affect the line emissivities
of highly ionized species \citep[e.g.,][]{gumf03}.  

We have developed a NEI spectral model that has the following characteristics:
(1) the ionization code is separated from the atomic data so that the latter can be updated
conveniently;
(2) the most updated atomic data are used; 
(3) recombinations into highly excited levels are included; 
(4) the electron temperature evolution, the dynamics,
and the ionization structure are determined self-consistently. 

\subsection{Atomic Data}
\label{sec-atomic}
Our atomic data are based on CHIANTI 
\citep[version 4.2]{youn03}, which provides a database of atomic energy levels,
transition wavelengths,
radiative transition probabilities (A-rates) and electron collisional excitation
rates for 23 elements from H ($Z=1$) to Zn ($Z=30$), totally 206 ions.  

Spectral calculations involving detailed fine structure levels are 
time-consuming and are not necessary for the current X-ray CCD spectral resolution. 
Therefore,  we have combined 
those fine structure levels together according to their degeneracies and have
re-calculated the collisional excitation rates and A-rates.
For example, for the hydrogen ion there are totally 25 energy levels listed in
CHIANTI.
After regrouping, we get six principle energy levels (1s, 2s, 2p, n=3, n=4 and n=5).
For n=3, there are five sub-levels 
($j=3s, ~3p_{1/2},~ 3p_{3/2}, ~3d_{3/2}, ~\rm and ~3d_{5/2}$),
which we regroup as one. The new collisional de-excitation rate and A-rate 
are $C_{3,i}=\sum_{j}\frac{g_{j}}{g}C_{3j, i}$ and
$A_{3,i}=\sum_{j}\frac{g_{j}}{g}A_{3j, i}$ respectively, 
where $g_{j}$ is the individual sub-level degeneracy and $g$ is the  total.
 
\begin{figure}[h]
   \centerline{
   \includegraphics[width=0.5\textwidth]{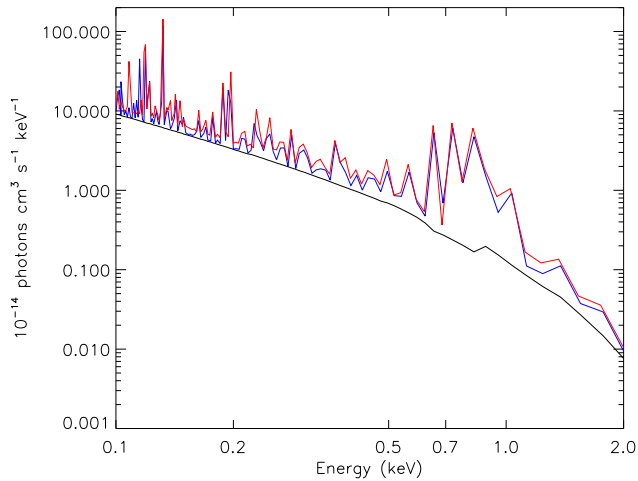}
       }
   \caption{ Comparison between the CIE spectra  of CHIANTI (blue) and our code
        with regrouped atomic data (red), assuming $T=5 \times 10^{6} \rm~ K$,
       cosmic abundance given by \citet{allen73},  and  the ion fractions from \citet{mazz98}.
        The black line denotes continuum. A spectrum bin size of 1\AA ~ is used.
       }
   \label{fig-com_cie}
  \end{figure}

A comparison between CHIANTI and our code with the
regrouped data is shown in Fig. \ref{fig-com_cie}. 
The total luminosity difference between the two spectra is less than 1\%. 
The difference in the line structure is very slight at $\gtrsim 0.5$ keV and 
somewhat more noticeable at lower energies. But such difference is well
within the uncertainties of the atomic data.
 % CIE comparison with CHIANTI and our code
 %figure 3

The ionization structure of a plasma is determined by the collisional ionization and radiative
and dielectronic recombinations \citep[e.g.,][and references therein]{mazz98}.  
CHIANTI  does not offer the rates of these atomic processes. We have taken the 
collisional ionization cross-sections from \citet{ar85}
and  \citet{ar92}, and radiative and dielectronic recombination rates from
\citet{vefe96} and \citet{mazz98},
except for Carbon, Nitrogen and Oxygen. For these three species, we
use the results of  \citet{naha00}, based on a large-scale close-coupling 
R-matrix calculation for photoionization cross-sections and recombination 
coefficients. 
We assume that three-body recombinations  are not important at the density 
range under consideration and neglect them for all ions. Our current code also 
does not include charge-exchange reactions. 

\begin{figure}[hb]
  \mbox{
  \includegraphics[height=3.5cm,width=7.7cm]{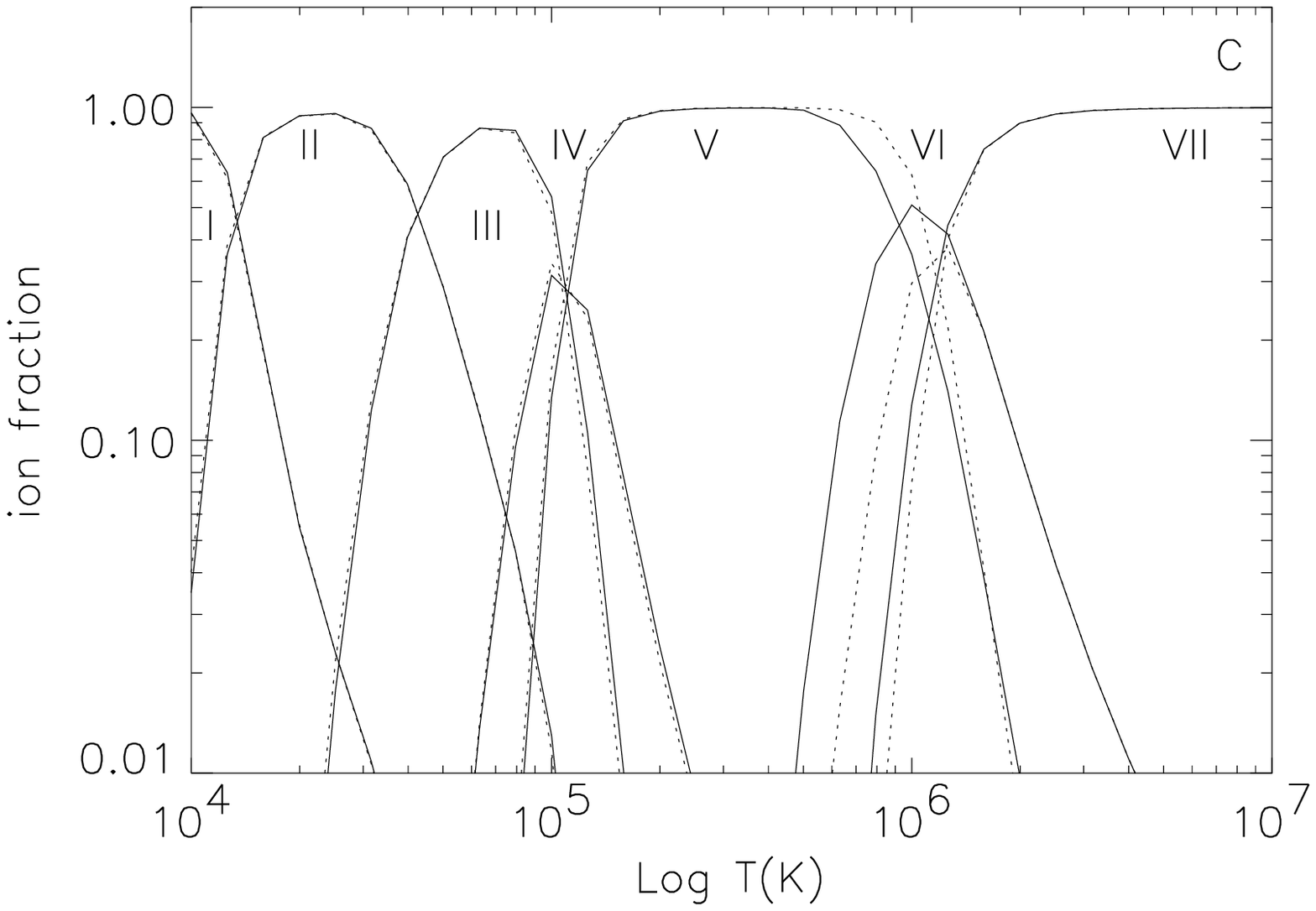}
  \includegraphics[height=3.5cm,width=7.7cm]{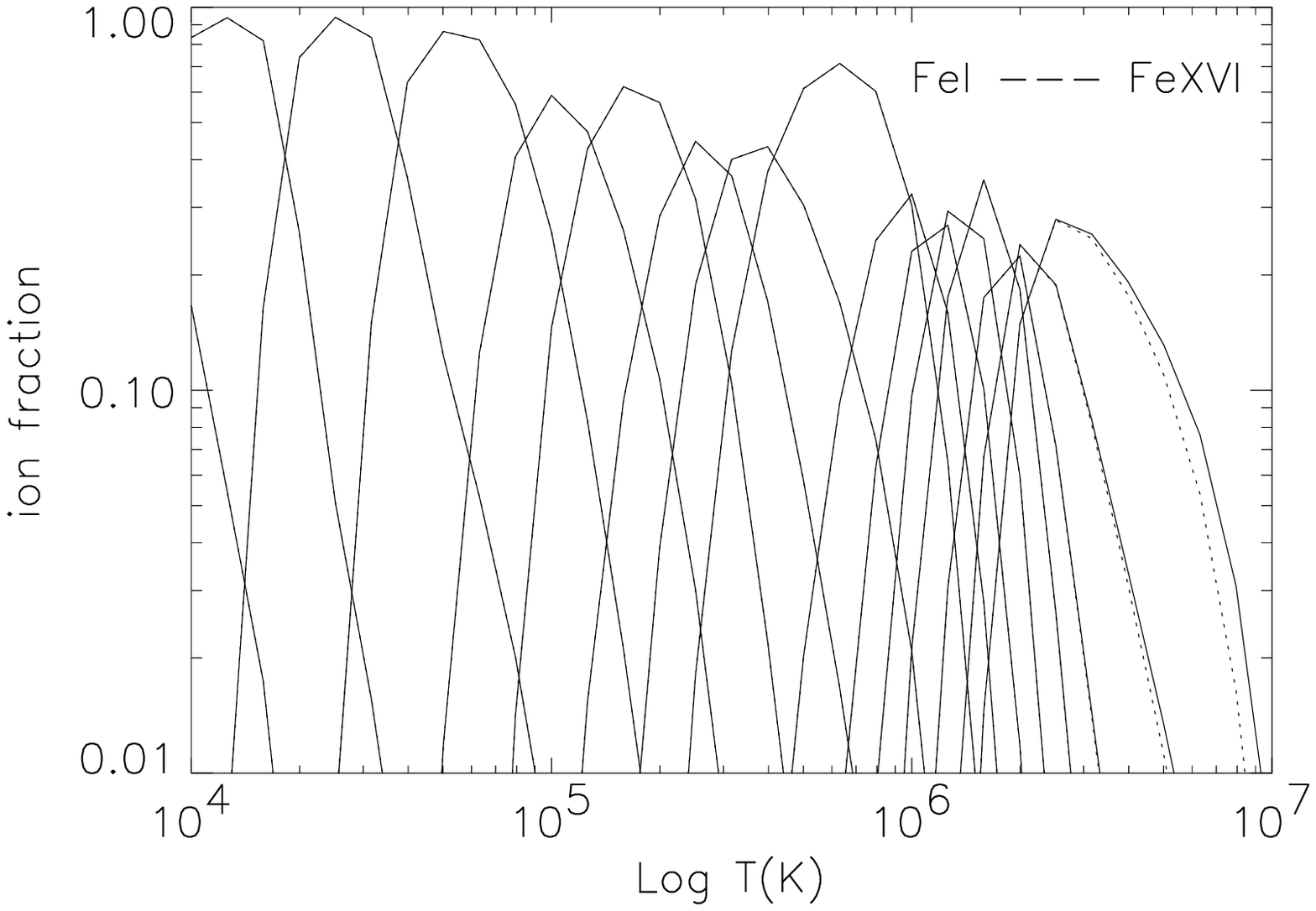}  
  }
  \mbox{
  \includegraphics[height=3.5cm,width=7.7cm]{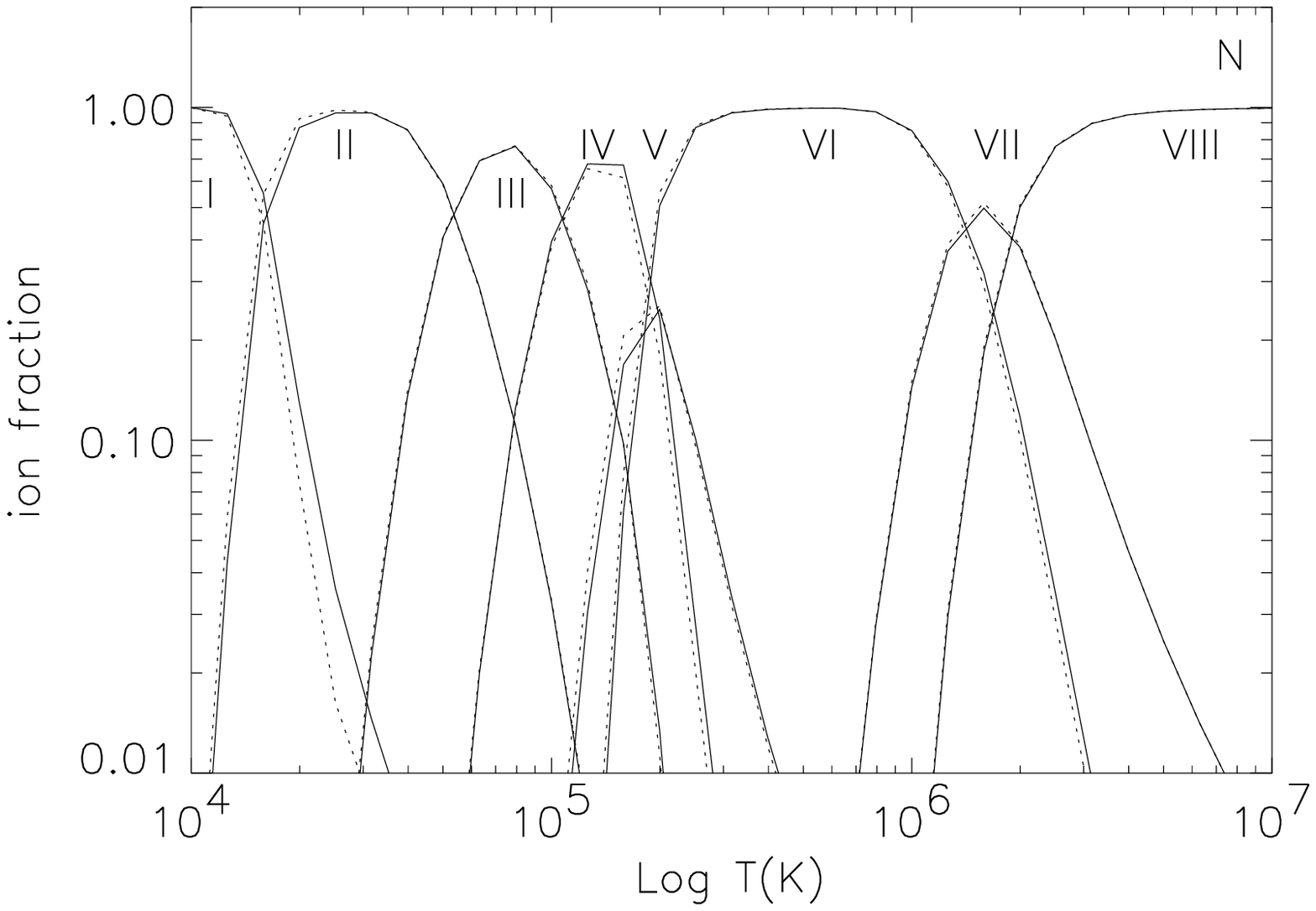}
  \includegraphics[height=3.5cm,width=7.5cm]{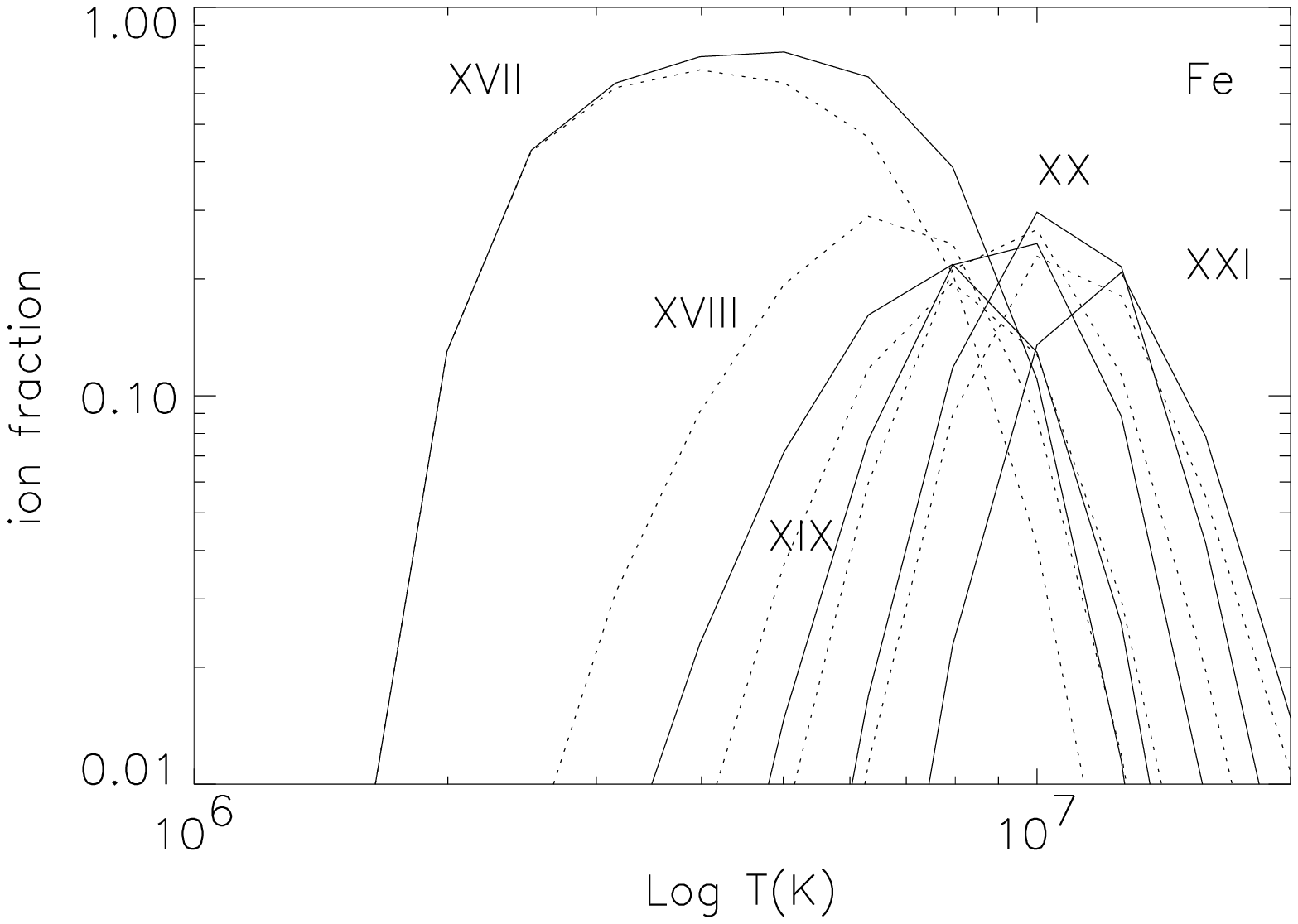}  
  }
  \mbox{
  \includegraphics[height=3.5cm,width=7.7cm]{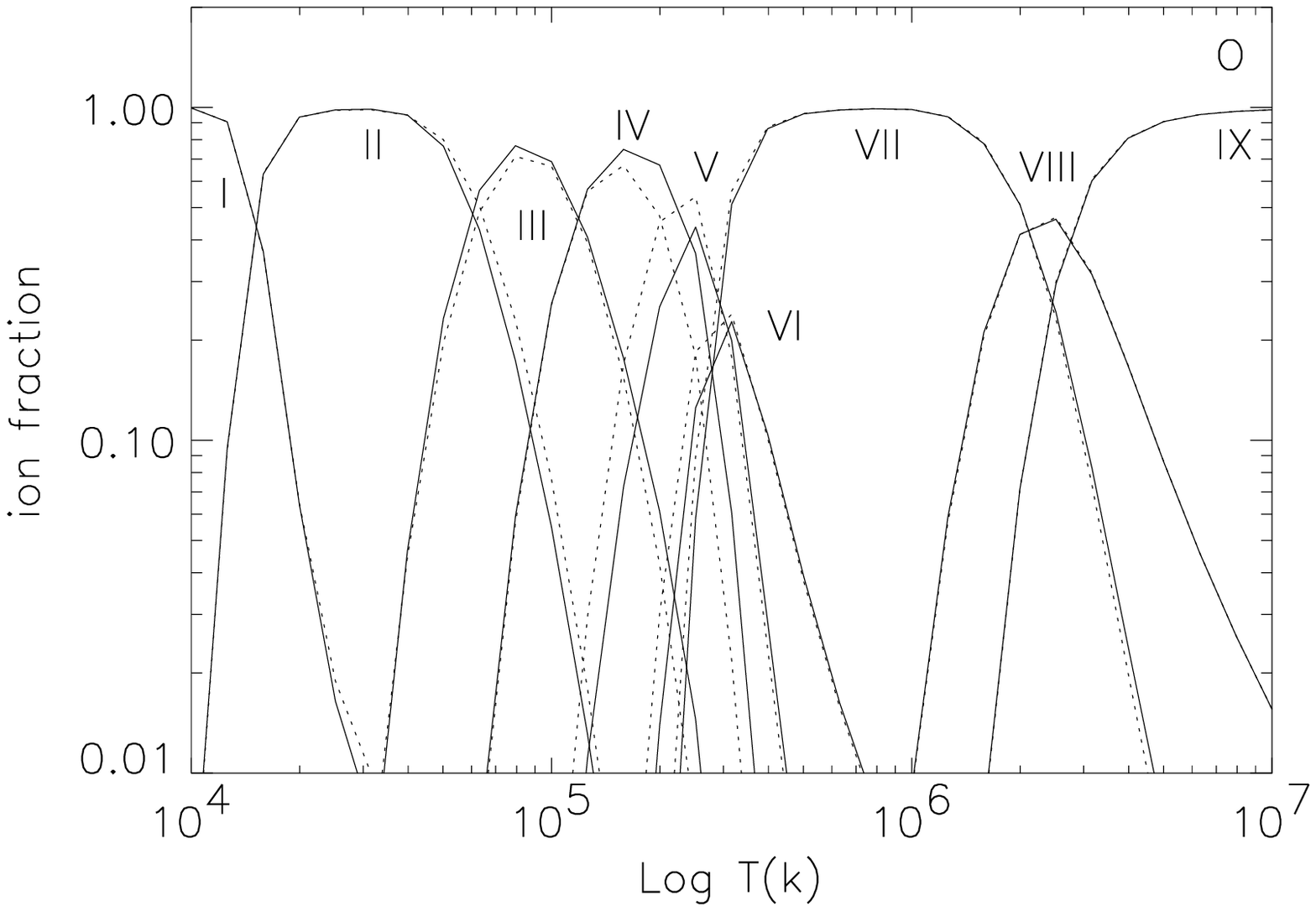}
  \includegraphics[height=3.5cm,width=7.7cm]{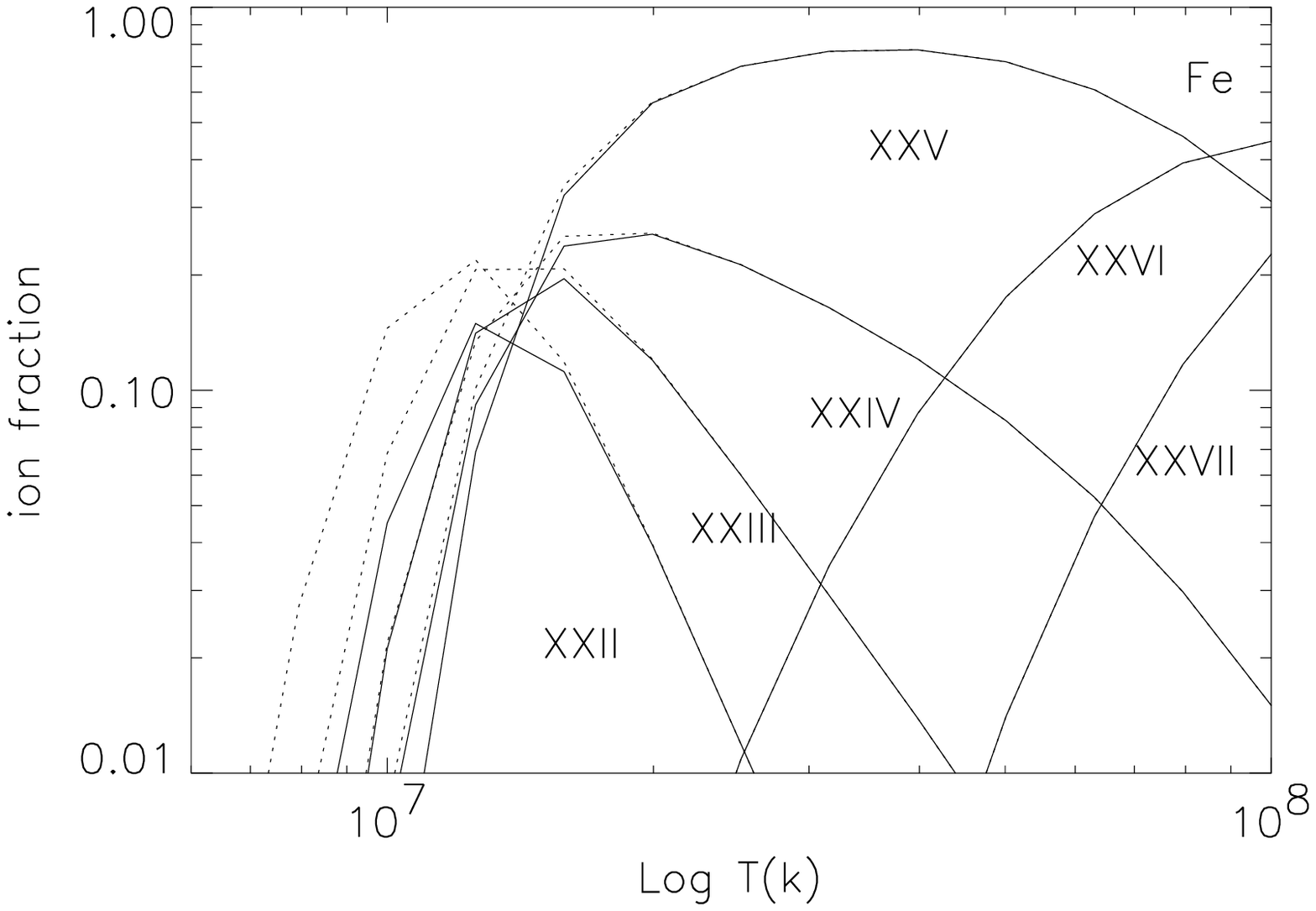}  
  }
  \caption{CIE ion fractions of our model (solid) and Mozz98's (dotted) for
        elements C, N, O, and Fe as marked in individual panels.
            }
  \label{fig-ion_cie}
  \end{figure}

Fig. \ref{fig-ion_cie} compares the CIE ionic fractions of our calculation with
those given by \citeauthor{mazz98} (1998, henceforth Mazz98). 
All, except for C, N, O and Fe, agree well because the same atomic data are used.
We find good agreement for all the ions of N,
with the difference being less than 10\%. For C and O, there is good agreement
except for CV, CVI, OIV, OV and OVI, for which the difference  can reach 
up to 50\% near the peak of maximum ionic fraction.  For example, 
at $T \approx 2\times 10^{5} \rm K$, our ionic fraction of OV is a factor 
of two less than that of Mazz98. From $T \approx 3\times 10^{6}\rm ~K$ 
to $T \approx 1.5\times 10^{7}\rm ~K$
where ionic fractions of Fe XV to Fe XXI are significant, our calculation is 
substantially different from Mazz98,  particularly for FeXVII to FeXXII. 
This is due to different radiative recombination rates employed.   
 %ionic fractions of our codes and comparisons   
 %figure 4

\subsection{Continuum and Line Emissions}
\label{sec-spectra}
The continuum emission consists of free-free, free-bound 
and two-photon continua. We include two important two-photon 
transitions: H-like $2s\rightarrow 1s $ and He-like  $ 1s2s
^{1}S_{0}\rightarrow 1s^{2}$ $^{1} S_{0}$. 

There are several ways to produce spectral 
lines in a plasma. Most spectral lines are produced by electron collisional excitation
of ions followed by the radiative decay of the excited level.
At X-ray energies free electrons can also excite an inner-shell 
electron to a level above the ionization potential.  
The excited level of the ion then undergoes either a radiative 
decay to a lower energy level to produce a satellite line, or an
autoionizing transition 
to the next ionization stage \citep{dere01}. We follow CHIANTI in treating 
this autoionizing transition as a radiative decay to the ground level  but
with no emission of radiation.
 In addition, we include the line emission  following an inner-shell
ionization. For example,
an impact ionization from the inner $1s^{2}$ 
shell of ion $Z^{+(z-1)}$ contributes to the production 
of a certain line in the next higher ion $Z^{+z}$.  We use the database of
\citet{mecr81}
for such lines (see \citet{gorc03} for an assessment of the database accuracy).

Line photons are also produced  during the cascade following a radiative
recombination into an excited level. This process is particularly important 
in  a delayed recombination scenario. A dielectronic
recombination leads to the capture of  an incident electron into an excited
state above the ionization threshhold of the recombined  ion and also 
emits a photon if an autoionization does not ensue.

\subsection{Level Balance Equations}
\label{level-pop}

Electron  populations at various energy levels of each ion are determined by
such processes as
collision excitation/de-excitation and radiative decay; the level balance
equations are:
  \begin{eqnarray}
  n_{i} \sum_{j \neq i} \alpha_{ij} = \sum_{j\ne i} n_j \alpha_{ji} \label{eqn-pop},
  \end{eqnarray}
  \noindent where $i$ and $j$ denote individual levels of an ion, 
$\alpha_{ji}$ is the total rate of the transitions 
between the two levels, and $n_i$ is the level $i$ population relative to
the  total level population. 

In a low density plasma, collisional excitation generally dominates 
over recombination in populating the excited states. 
In an over-cooled NEI state, however, the temperature may be sufficiently low that 
recombination can make a non-negligible contribution to the level population
\citep[e.g.,][]{dere01}.
We include radiative recombinations into the principle energy levels ($n=1 $ to 6)   
of each ion; the level balance equation 
for ion $Z^{+z}$ (element $Z$ with $z$ electrons removed) is
  \begin{eqnarray}
   n_{i} \sum_{j \neq i} \alpha_{ij} = \sum_{j\ne i} n_j \alpha_{ji}+n_{\rm
e}\frac{N(Z^{+(z+1)})}{N(Z^{+z})}RR(Z^{+(z+1)})_{i} 
   \end{eqnarray}
   \begin{eqnarray}
   \hspace*{1.1in}=n_{j}(n_{\rm e} C_{ji} + A_{ji}) + n_{\rm
e}\frac{N(Z^{+(z+1)})}{N(Z^{+z})}RR(Z^{+(z+1)})_{i} \label{eqn-poprr}
  \end{eqnarray}
where $C_{ji}$ is the electron collisional rate, 
$A_{ji}$ is the radiative decay rate (zero if $j < i$), and  
$RR(Z^{+(z+1)})_{i}$ is the radiative recombination rate 
to the level $i$ of ion $Z^{+(z)}$. Following \citet{oster89}, we calculate
the rate using Milne's relation and the photoionization cross sections
of \citet{kala61} for excited
states and those of \citet{veya95} for ground states.  Thus,
  \begin{eqnarray}
   RR(Z^{+(z+1)})_{i}=\int_{\nu}\alpha_{\nu,i}d\nu,
  \end{eqnarray}
 and 
  \begin{eqnarray}
  \alpha_{\nu, i}
=\frac{4}{\sqrt{\pi}}\frac{g_{i}}{g_{z+1}}\frac{h^{3}\nu^{2}}{c^{2}(2m_{e}kT)^{3/2}}e^{-h(\nu-\nu_{T})/kT}\sigma_{phot, i}(\nu)
  \end{eqnarray}
where $g_{i}$ and $\sigma_{phot, i}$ are the degeneracy  and  
photoionization cross section of level $i$ of ion $Z^{+(z)}$.

\subsection{Non-equilibrium Ionization Calculations}
\label{sec-ioniz}
In general the dynamics and ionization of a plasma 
are entangled. The ionization structure, through its heating/cooling effects, 
can affect the temperature evolution. However, in a cluster wind, 
radiative cooling can typically be 
neglected and the dynamics and ionization of the wind 
can be separated. With the mass input having a certain 
spatial distribution in our model, and assuming the newly injected gas to be neutral, 
the cluster wind number density of the neutral ion $N_{Z,0}$ for a given 
element $Z$ evolves in a cluster wind as:
    \begin{eqnarray}
     \frac{1}{r^{2}}\frac{dN_{Z,0}ur^{2}}{dr} =\frac{{\rm Z}\dot{m}(r)}{m_{Z}}
-n_{e}N_{Z,0}S_{Z,0}+n_{e}N_{Z,1} \alpha _{Z,1} \label{eqn-is2},
    \end{eqnarray}
with Z and $m_{Z}$ being the abundance and mass of element $Z$.

 The number density of other ions ($N_{Z,z>0}$) of the same element will follow:
    \begin{eqnarray}
     \frac{1}{n_{e}}\frac{dN_{Z,z>0}}{dt} = N_{Z,z-1}S_{Z,z-1}
-N_{Z,z}(S_{Z,z}+\alpha _{Z,z})+N_{Z,z+1} \alpha _{Z,z+1}, \label{enq-is3}
    \end{eqnarray}
where $S_{Z,z}$ and $\alpha_{Z,z}$ are the total ionization ($z\rightarrow z+1$)
and recombination ($z\rightarrow z-1$) rate coefficients (in $\rm
cm^{3}~s^{-1}$) of 
ion $Z^{+z}$. Our interest here is in highly ionized ions. Therefore,
the exact ionization state of the injected gas is not important as long as 
it is cold or warm (i.e. $T\lesssim 10^{4}~\rm K$).

For each element, we simultaneously solve these differential equations 
using IDL  code LSODE which adaptively solves
stiff and non-stiff systems of ordinary differential equations. 

In an evolving plasma, a characteristic time scale to approach 
a CIE state is  $\tau_{CIE} \approx [n_{e}(S_{Z,z}+\alpha_{Z,z})]^{-1} $ 
\citep{mewe97}. The CIE approximation requires the cooling/heating time be 
much longer than $\tau_{CIE}$. 
As mentioned earlier, CIE is only a special case. 
If experiencing a rapid heating or cooling 
process such as thermal instability, shock, or rapid expansion,
a plasma may be under-ionized 
or over-ionized, compared to the CIE result \citep{dopi02}. 

\begin{figure}[h]
    \centerline{
   \includegraphics[width=0.5\textwidth]{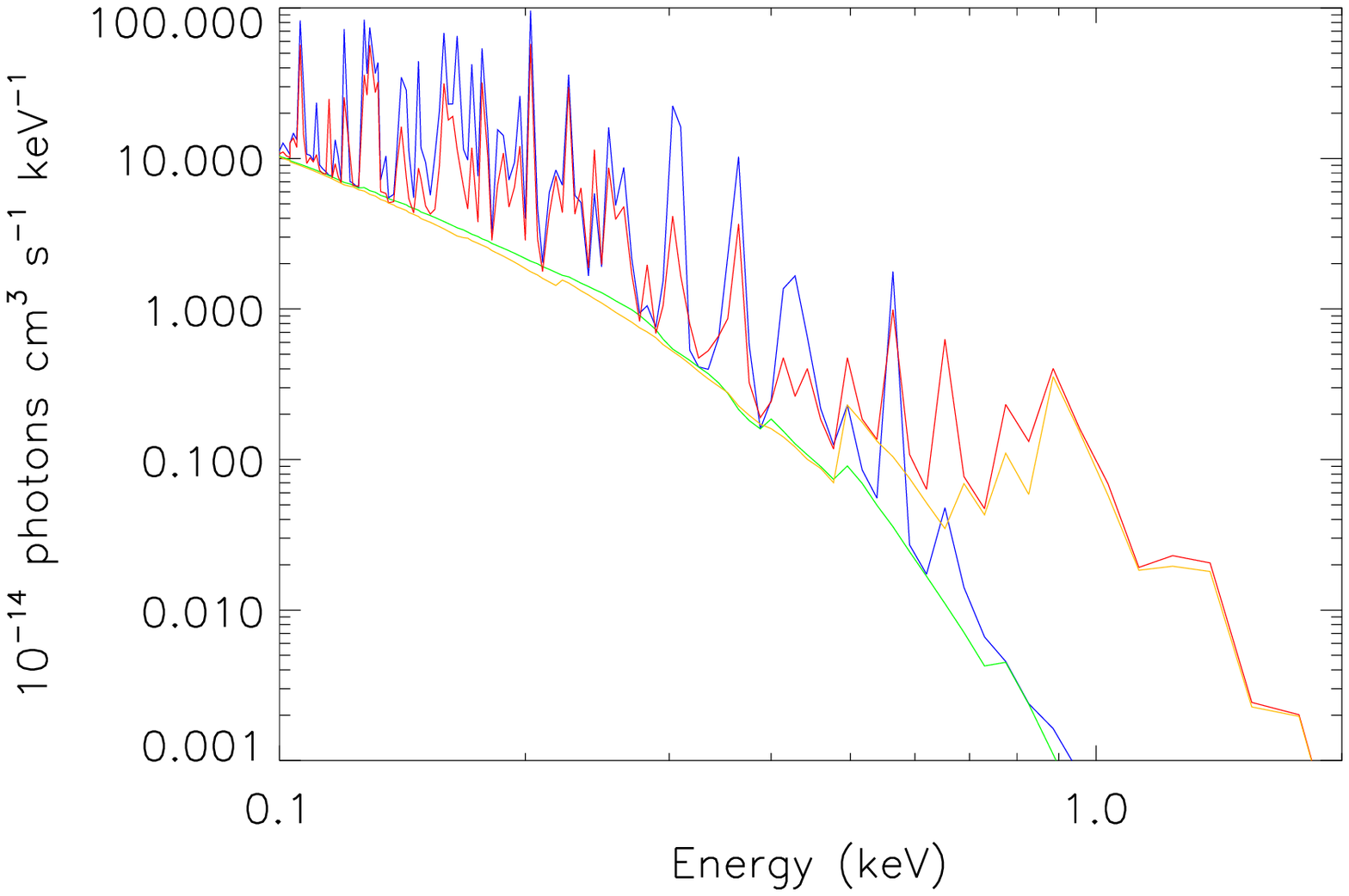}
   \includegraphics[width=0.5\textwidth]{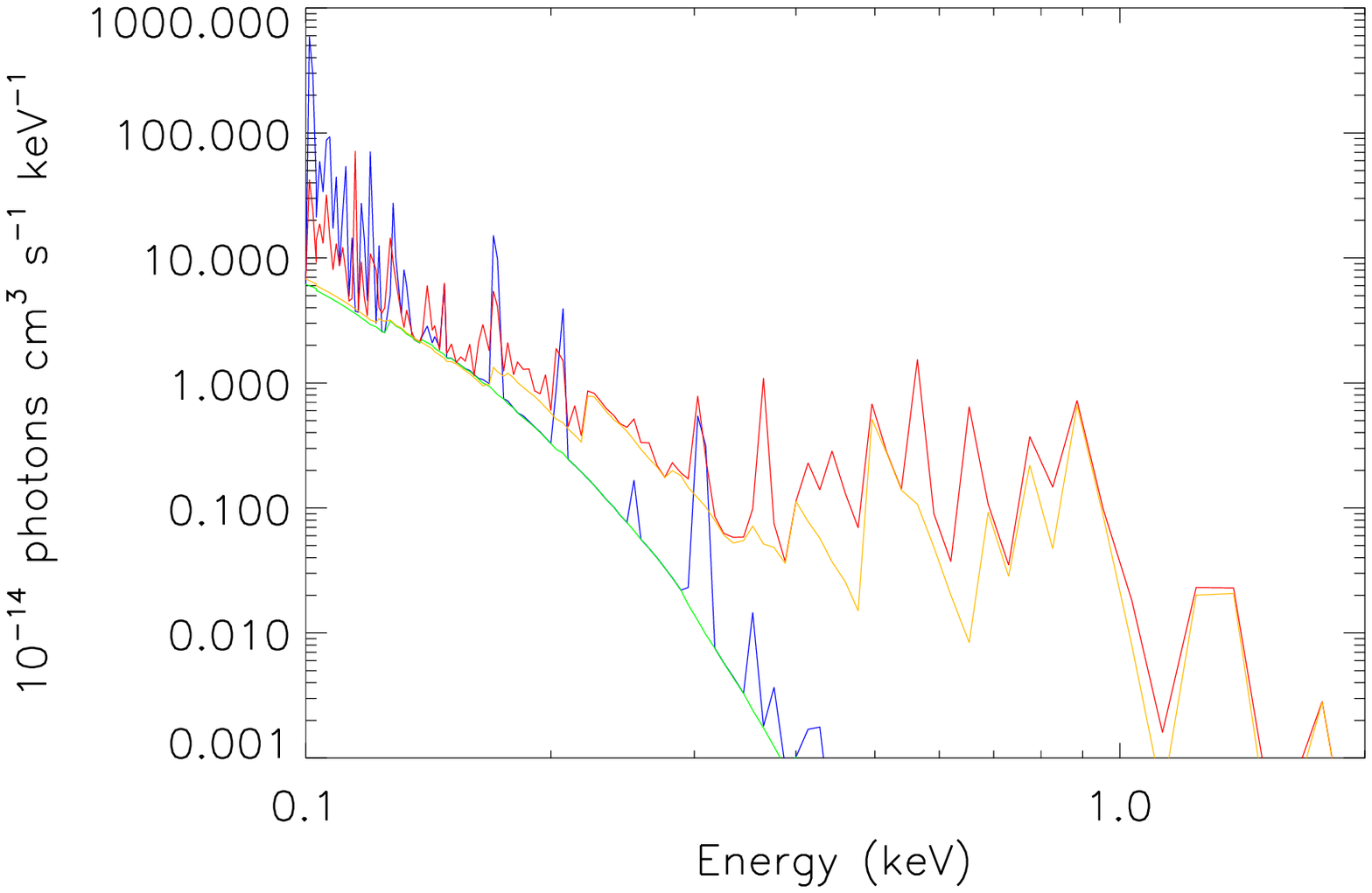} 
       }
   \caption{ Spectra of an adiabatically expanding
   cluster wind at $r = 1$ pc (left panel) and 2 pc (right panel): CIE (total --- 
   blue line; continuum only --- green line ) and non-CIE (total --- red
   line; continuum only --- yellow line) calculations.
       }
   \label{fig-com_n2}
   \end{figure}

\begin{figure}[h]
    \centerline{
   \includegraphics[width=0.5\textwidth]{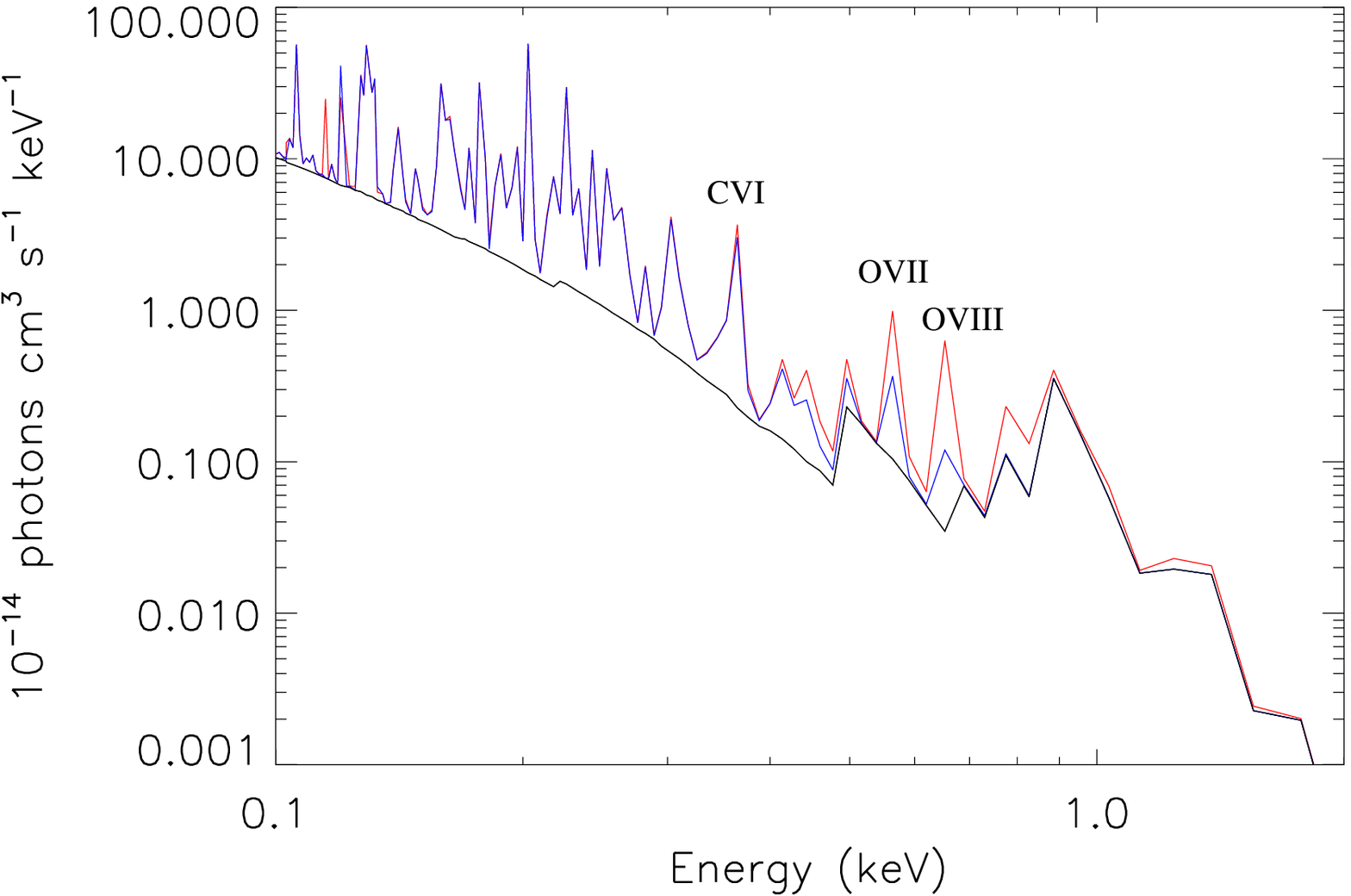}
   \includegraphics[width=0.5\textwidth]{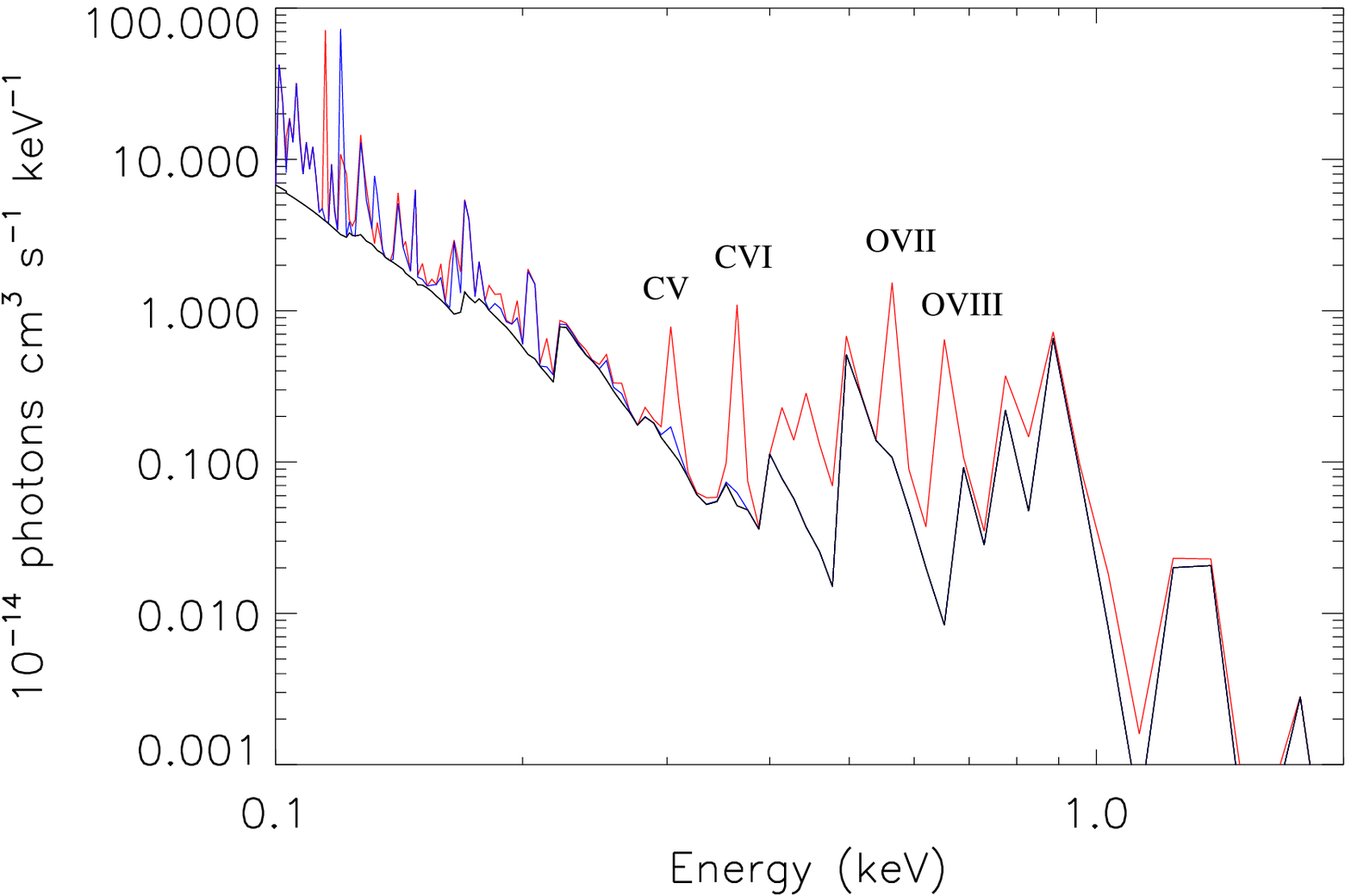} 
       }
   \caption{ Spectra with (red) and without (blue) adding the cascades following 
                    radiative recombinations to highly
                   excited levels for an adiabatically expanding stellar 
                    cluster wind (see text for details). Black line denotes continuum only.  
       }
   \label{fig-com_addrecom}
   \end{figure}

As an illustration, we consider a toy model for 
an adiabatically expanding stellar cluster wind
with a constant mass input rate 
$\dot M\sim 3\times 10^{-5}\rm ~ M_{\odot}~yr^{-1}$ and a constant velocity 
$v=1000 \rm ~km s^{-1}$. We assume that the wind is injected at an initial 
radius of $r_{0}=0.3$ pc and is  heated up to a CIE state
with an equilibrium temperature $T_{0}=5\times 10^{6} \rm ~K$. As the wind
expands adiabatically, 
its temperature drops as $T=T_{0}(r/r_{0})^{-4/3}$.  Fig. \ref{fig-com_n2} 
compares the CIE and non-CIE spectra at
$r=1~{\rm pc, where}~T \sim 1\times 10^{6} \rm~ K$ (left panel)
and at $r=2~{\rm pc, where}~ T \sim 4\times 10^{5} \rm ~ K$ (right panel).
The difference is significant: in the lower energy part, 
the line intensities in the non-CIE spectrum are weaker
than those in the CIE case, while in the upper energy part saw-tooth
structures are present in the non-CIE spectra; all these are due to delayed recombinations. 
% simple expanding model to test our code
% figure 5    
Fig. \ref{fig-com_addrecom} shows the spectra at the same two radii with and 
without including the cascade of electrons following recombinations to $n > 1$ 
levels. The cascade lines of CV, CVI, OVII and OVIII are clearly visible.
% adding RR effect
%figure 6

\begin{figure}[h]
   \centerline{
   \includegraphics[width=0.65\textwidth,angle=90]{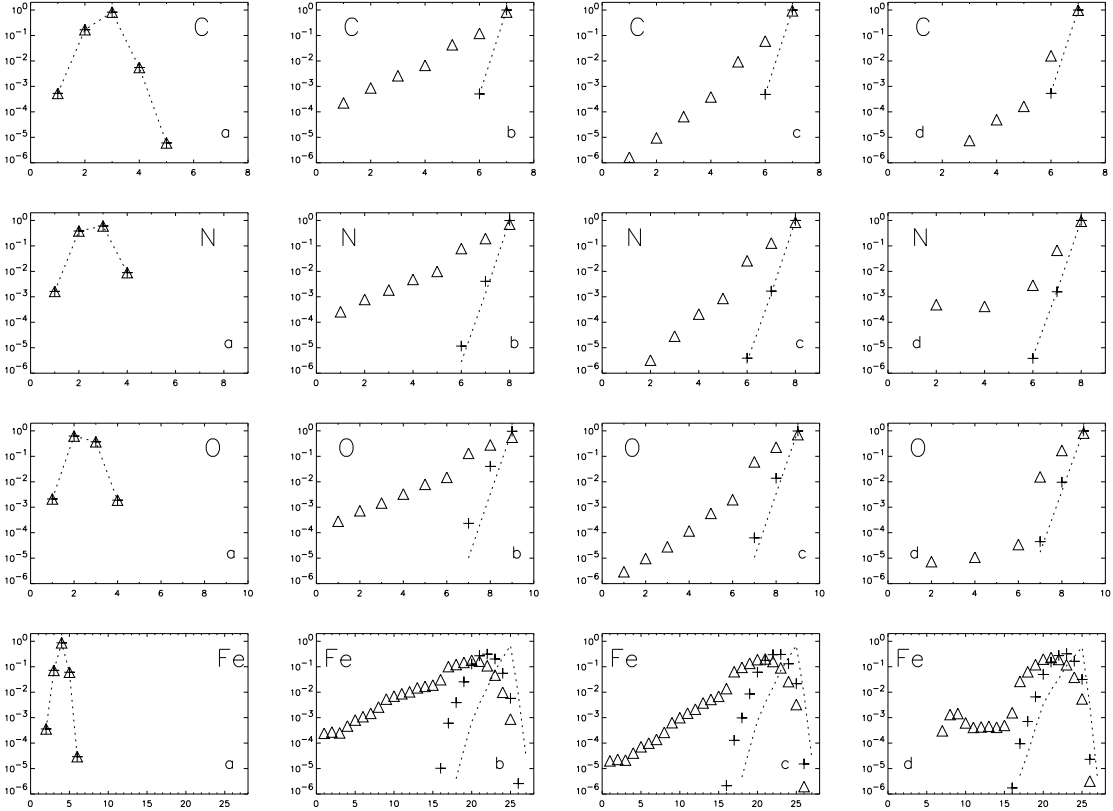}
   }
   \caption{C, N, O and Fe ion fractions calculated for a cluster wind 
           at four radii: panel a ($0.003 \rm~ pc$), $b$
($0.30 \rm ~pc$), 
	    $c$ ($0.72 \rm ~ pc$) and $d$ ($1.38 \rm ~pc$). 
            Two mass injection scenarios are assumed: all at 
	    the center (crosses,  and the dotted lines denote the 
            corresponding CIE calculation) and 
            in an exponential distribution (triangle). 
	    Y-axis denotes the ion fraction, X-axis denotes
	    the ionic stage (e.g., 1 denotes I).     
    }
 \label{fig-neion_m}
\end{figure}

The more general model for a cluster wind with distributed mass and energy
injections has been described in \S 2.1, and the 
radial profiles of the wind are shown in Fig. \ref{fig-dyn}. 
We use the case of $\dot{M}_{0}=2.3 \times 10^{-4}~\rm M_{\odot} ~yr^{-1}$,
 $V_{\infty}= 2000 \rm~ km~ s^{-1}$, and a sonic radius $r_{s}=1.38$ pc to illustrate
 the effect of a distributed mass injection on the ionization structure.
Fig. \ref{fig-neion_m} 
compares the ionic fractions of C, N, O and Fe at four radii between the case in which $\dot{M}_{0}$ 
is all injected at the center and that in which it is injected in an exponential distribution. 
 The effect of the distributed
mass injection clearly shows up as a long ionization tail. 
%the distribution of ion stages.
% mass injection effect 
%figure 7   

\section{MODEL PREDICTIONS FOR STELLAR CLUSTER WINDS}
\label{sec-predictions}
\begin{figure}[h]
   \centerline{
   \includegraphics[height=1.0\textwidth,angle=90]{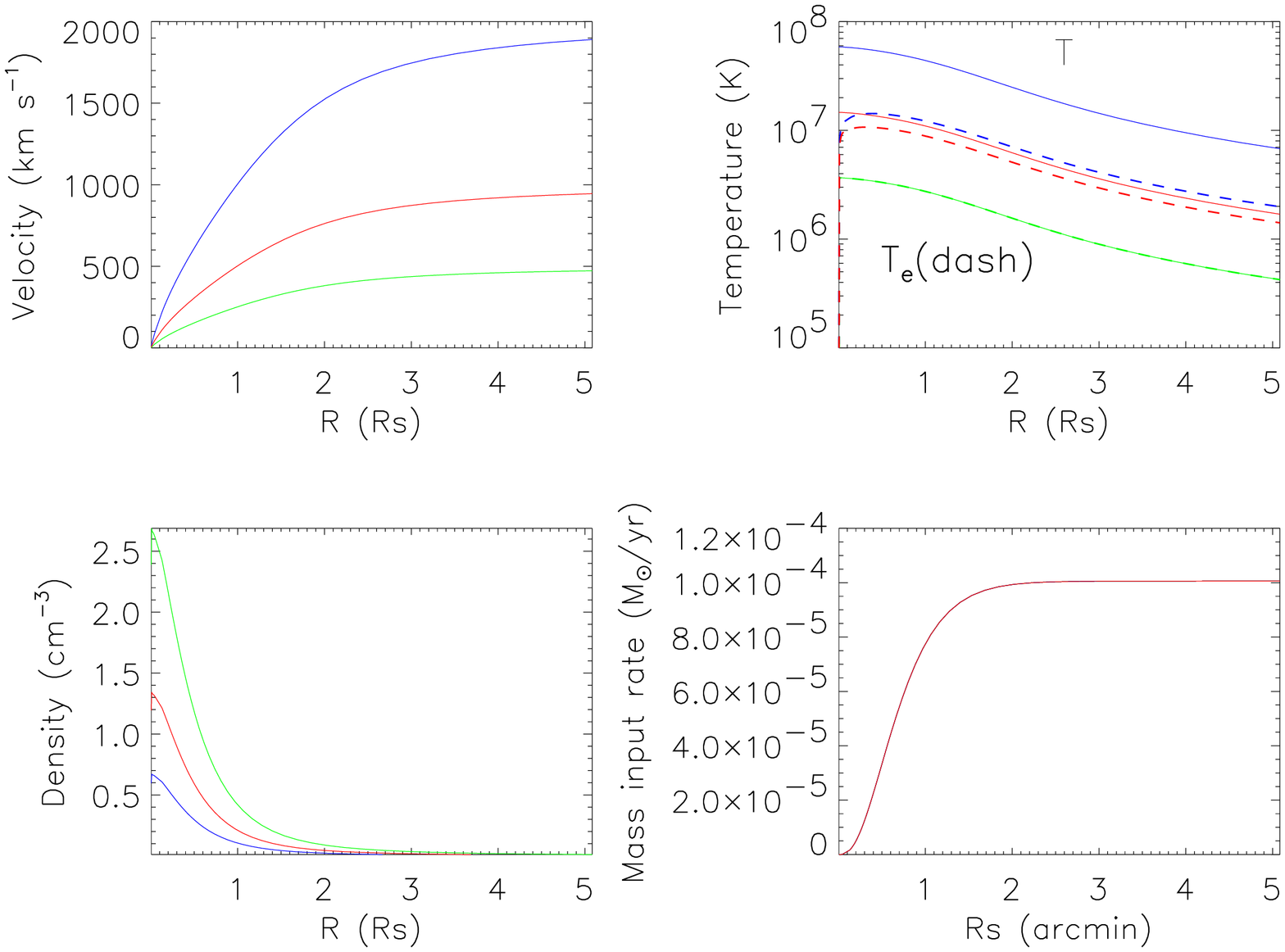}
    }
 \caption{ Cluster wind profiles for:    
  $V_{\infty}=500~ (blue),  ~1000 ~(red), ~ \rm and ~2000 ~(green) ~\rm km~s^{-1}$. 
  The mass injection rate is the same ($\dot {M}_{0}=10^{-4} \rm ~M_{\odot}~yr^{-1}$). 
  The non-equilibration electron temperature ($T_{e}$) in each case is plotted separately (dash lines).}
  \label{fig-grid-m4-struc}
\end{figure}

To illustrate  the condition under which the NEI effect is significant for stellar 
cluster winds, we here consider a cluster of an exponential stellar 
distribution with a scale radius $r_{sc}=\rm 0.48~pc$,
corresponding to a sonic radius at $R_{s}=1.97\rm~pc$.
We calculate results for a parameter grid: 
$\dot{M}_{0}=10^{-4} ~\rm and ~ 10^{-3}~\rm M_{\odot} 
~yr^{-1}$; $V_{\infty}=500, 1000, \rm ~and~ 2000 \rm~ km~ s^{-1}$. 
 The velocity, temperature, and density profiles of
the wind are illustrated in Fig. \ref{fig-grid-m4-struc}.  
We consider both the equilibration and non-equilibration
cases of the electron and ion temperatures (\S 2.1)
and both the CIE and non-CIE scenarios for comparison. 
To demonstrate the results,
we calculate the intrinsic cumulative spectra for three annular regions, 
[0, $R_{s}$], [$R_{s}$, 2$R_{s}$], and [2$R_{s}$, 4$R_{s}$], as well as 
the predicted {\it Chandra} ASIS-S count intensity distribution 
over the [0, 2$R_{s}$] range and in two
energy bands, [0.3, 2.0] keV and [2.0, 8.0] keV. For illustration, we assume
that the distance to the cluster is $D = 10$ kpc and that the metal
abundance is solar.

%wind profiles for theoretical investigation
%figure 8

\begin{figure}[h]
   \centering
   \mbox{
   \hspace*{-0.2in}
  \includegraphics[height=0.5\textwidth,width=0.4\textwidth,angle=90]{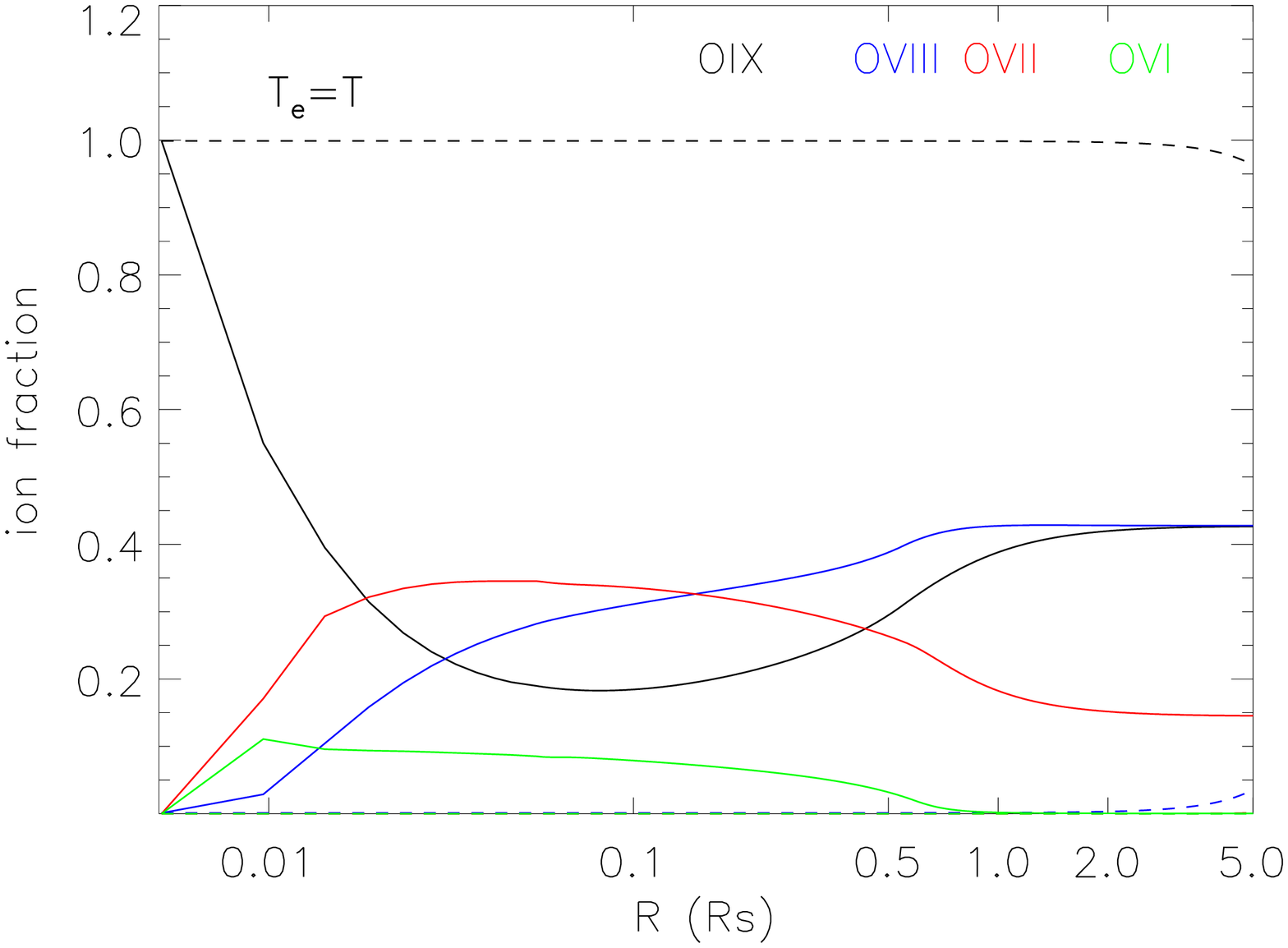}
  \includegraphics[height=0.5\textwidth,width=0.4\textwidth,angle=90]{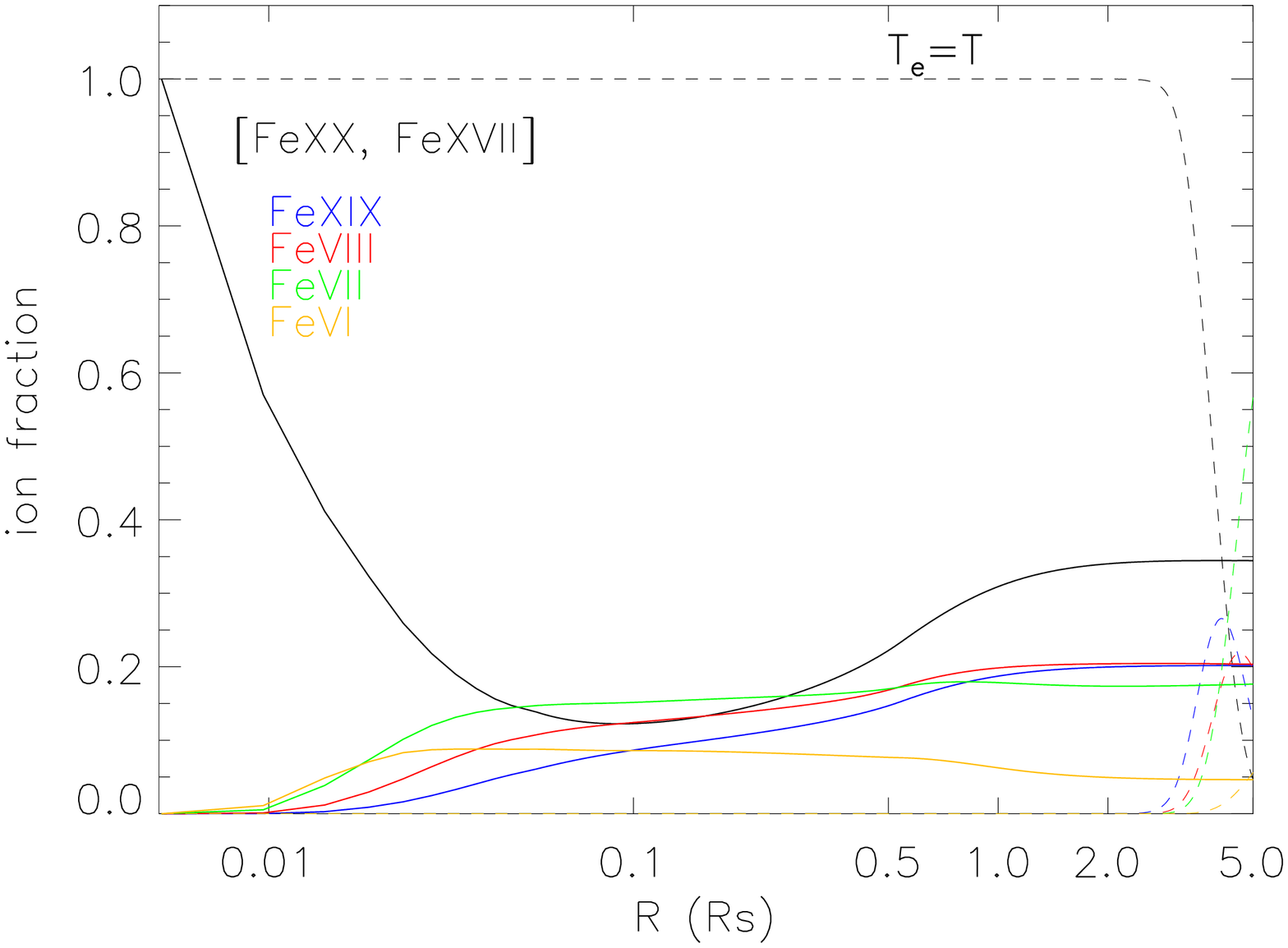}
   }
   \vskip 0.2in
   \mbox{
   \hspace*{-0.2in}
  \includegraphics[height=0.5\textwidth,width=0.4\textwidth,angle=90]{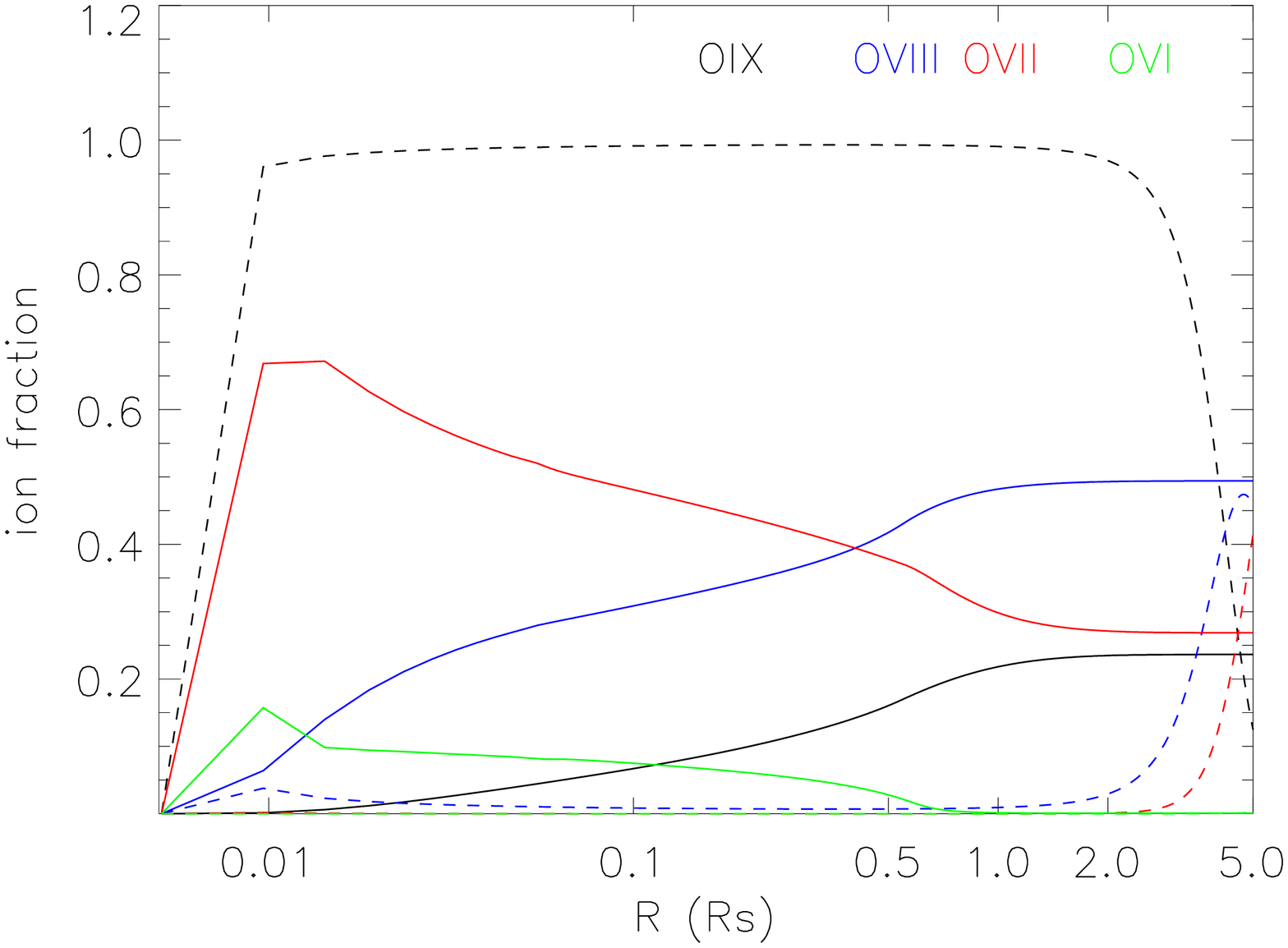}
  \includegraphics[height=0.5\textwidth,width=0.4\textwidth,angle=90]{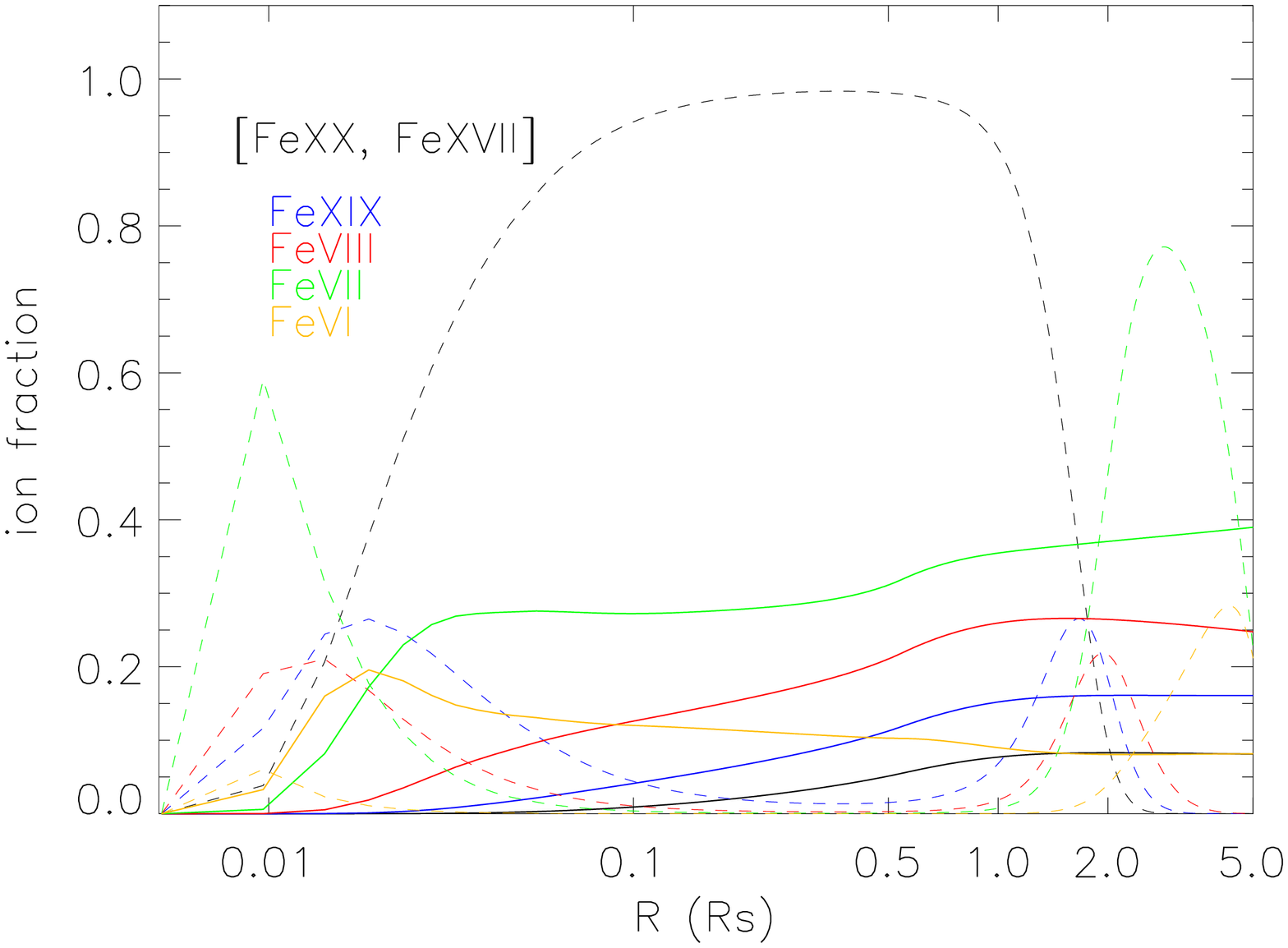}
   }
   \caption{Comparison of the ion fractions between 
            the CIE (dash) and the non-CIE (solid) calculations  of the cluster wind with
            $\dot{M}_{0}=1\times 10^{-4} ~\rm M_{\odot}~yr^{-1} $ \rm and ~
            $V_{\infty}=2000~\rm km~s^{-1}$
             for $T_{e}=T$ (upper row) and $T_{e}$ calculated via 
	     Coulomb interaction (bottom row). 
               }  
  \label{fig-neion_ionf_com}
\end{figure}

First, we find that the separate $T_{e}$ evolution does not lead to 
significant spectral difference in the cumulative spectrum. It is already seen from Fig.
\ref{fig-grid-m4-struc}
that $T_{e}$ is significantly lower than $T$ only when the density is low (i.e. 
$V_{\infty}=2000 ~\rm km~s^{-1}$ and $\dot {M}_{0}=10^{-4} ~\rm M_{\odot}~ yr^{-1}$).
Even in this case the ionic fractions of the key O and Fe ions are fairly
similar (Fig. \ref {fig-neion_ionf_com}). 
The biggest difference in the ionic fractions between the two $T_{e}$ evolutions
 occurs during the
rise of $T_{e}$ from $10^{5}$ K to $10^{7}$ K in the non-equilibration evolution
($R \lesssim 0.01 R_{s}$).
In the region making the most contribution to the X-ray flux 
($2R_{s}\gtrsim R \gtrsim 0.05 R_{s}$),  the ionic
fractions are comparable between the two $T_{e}$ evolutions 
in either the CIE or non-CIE case. 
Henceforth in the following discussions on the cumulative 
spectrum and count intensity distribution, only 
the results from the $T_{e}$ evolution via Coulomb interactions are used.
%
%comparison with/without coulomb interaction
%figure 9

\begin{figure}[h]
  \centering
  \includegraphics[height=1.0\textwidth,angle=90]{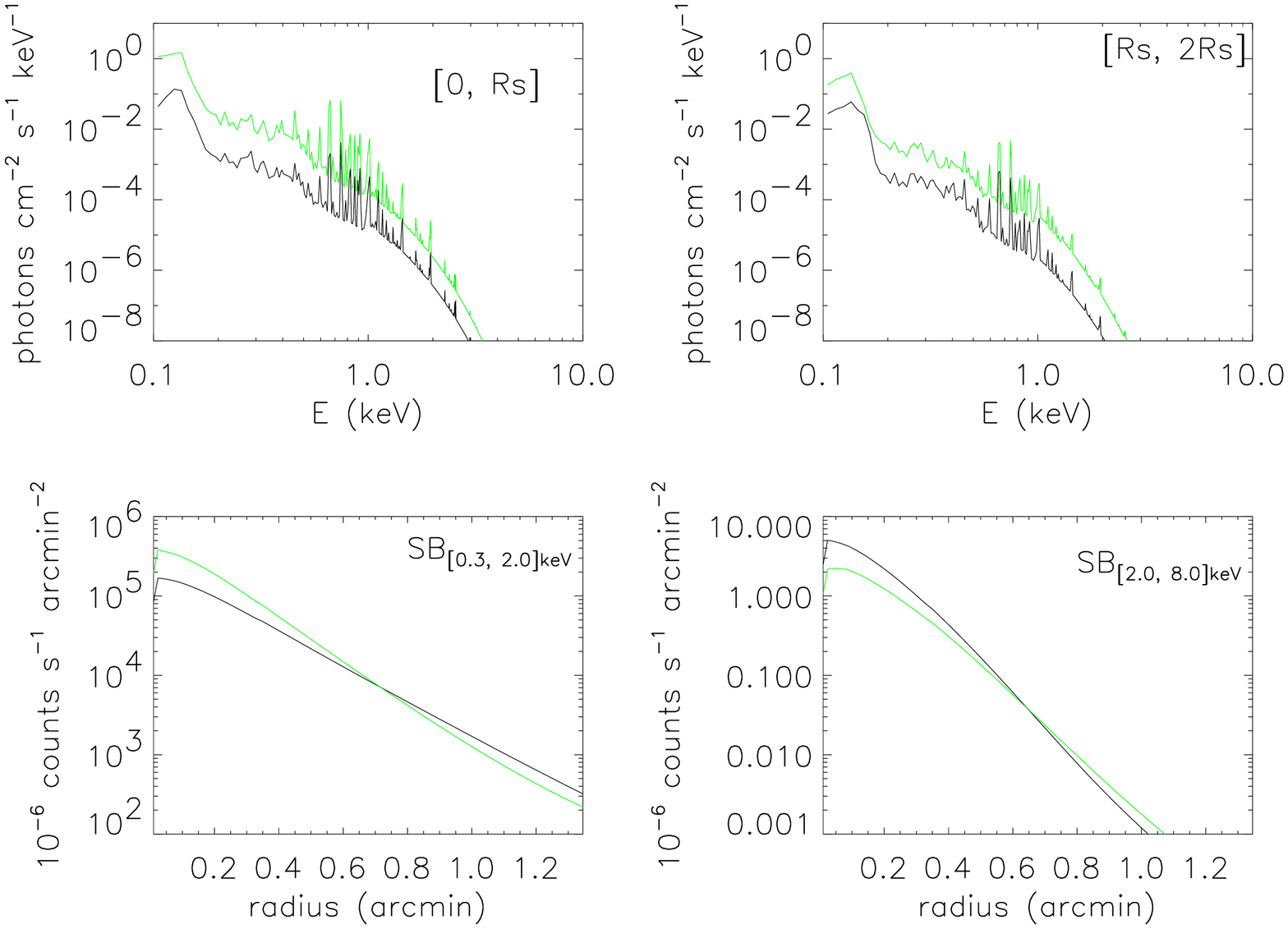}
 \caption{Non-CIE model predictions (green), compared with CIE calculations
               (black) for the wind profile with  $V_{\infty}=500 ~\rm km~s^{-1}$, as 
	        illustrated in Fig. \ref{fig-grid-m4-struc}. 
	       Upper panel: Cumulative spectra within annuli [0, $R_{s}$] (right) and [$R_{s}$, 2$R_{s}$] (left);
	       for clarity, each cumulative spectrum from non-CIE calculations has been increased by a factor of 10.
	       Lower panel:  {\it Chandra} ACIS-S count intensity distributions in [0.3, 2.0]~keV (right) and [2.0,
               8.0]~keV (left) bands. 
               }
   \label{fig-grid-m4-v5}
 \end{figure}

\begin{figure}[h]
   \centering
 \includegraphics[height=1.0\textwidth,angle=90]{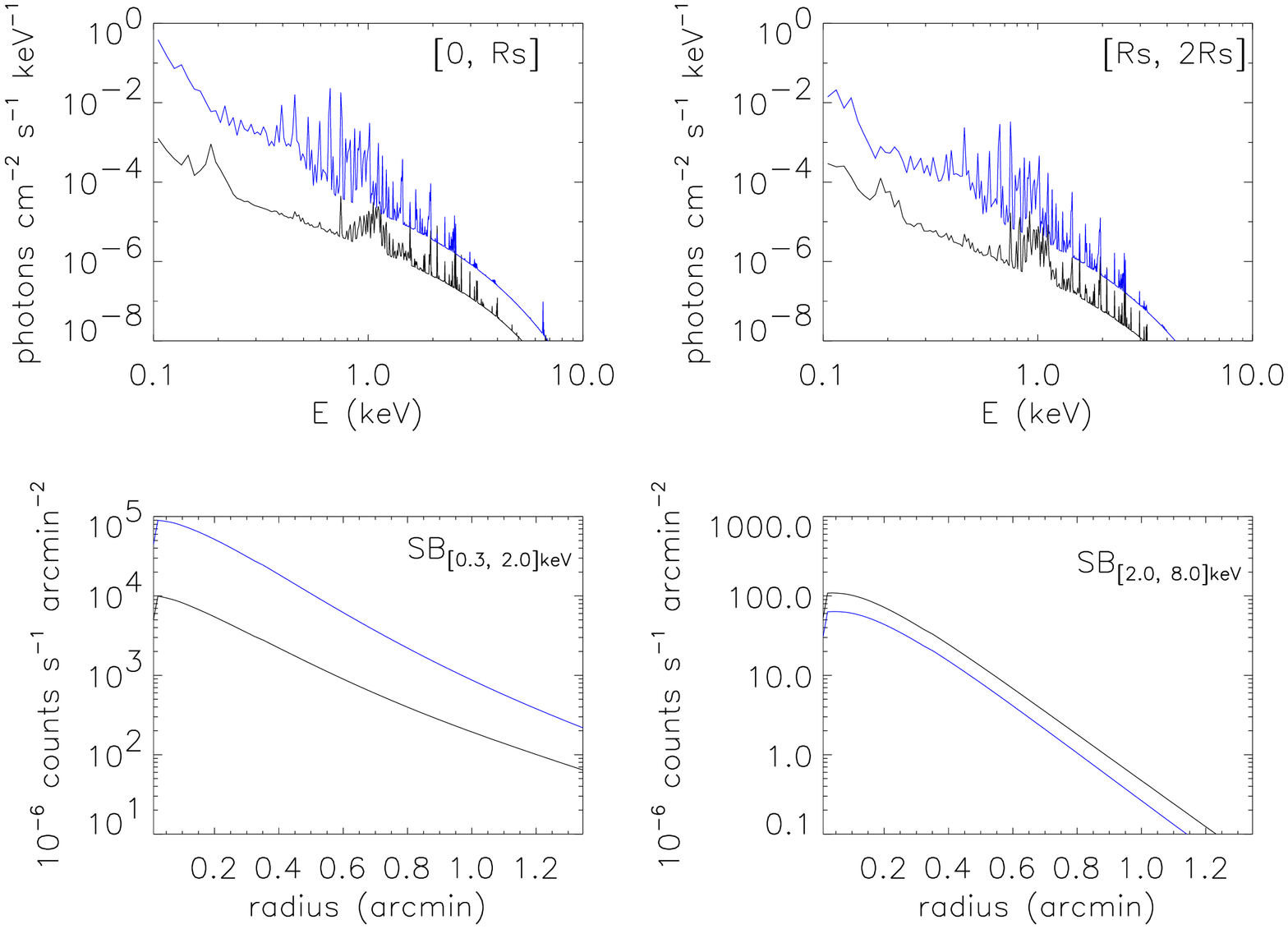} 
 \caption{Same as Fig. \ref{fig-grid-m4-v5} but for the  
               wind profile with $V_{\infty}=2000 ~\rm km~s^{-1}$ , as illustrated in Fig.
\ref{fig-grid-m4-struc}.  }
   \label{fig-grid-m4-v2}
 \end{figure}

\begin{figure}[h]
   \centering
  \mbox{
  \hspace{0.1in}
  \includegraphics[height=0.47\textwidth,angle=90]{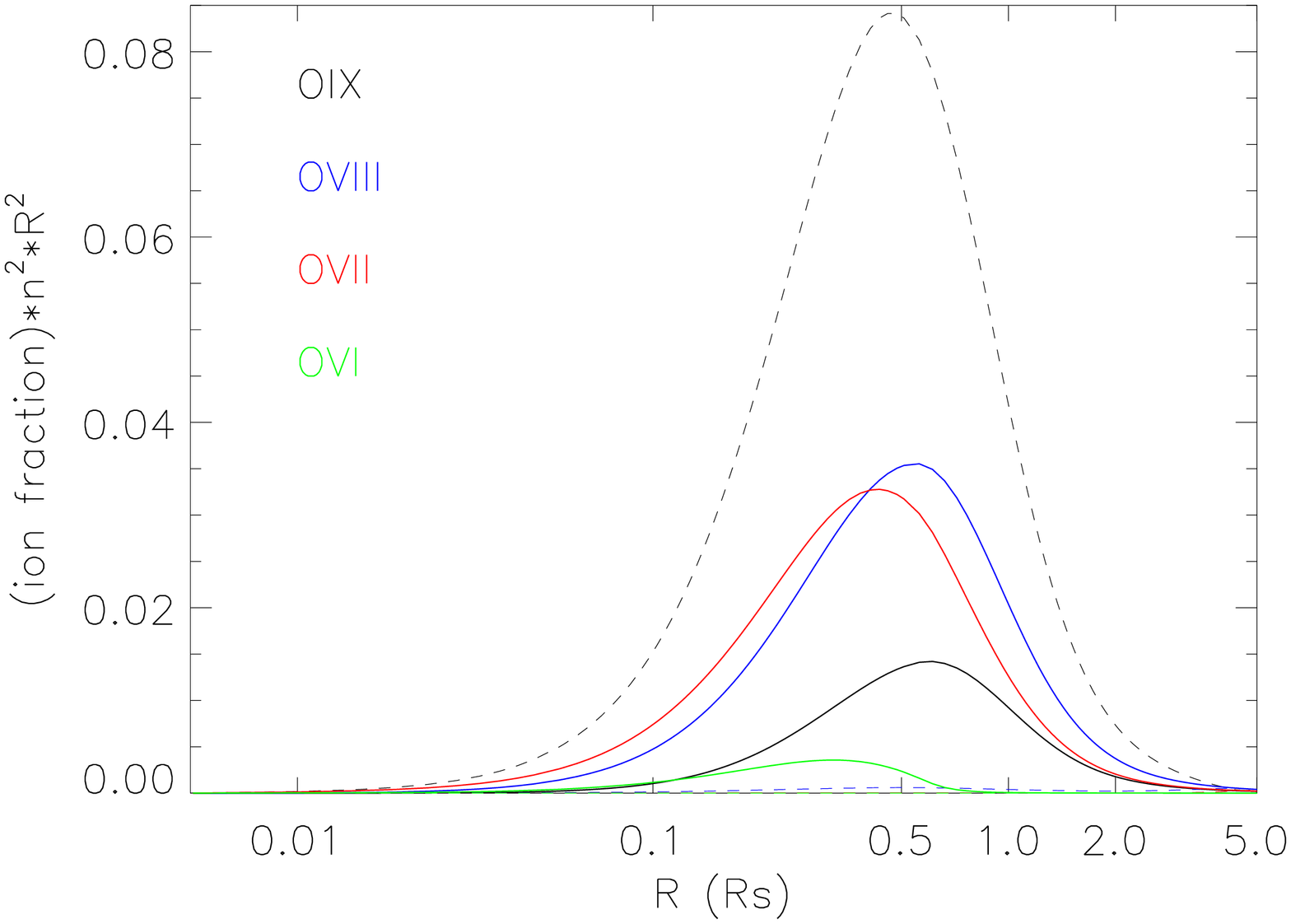}
   \hspace{0.16in}
  \includegraphics[height=0.47\textwidth,angle=90]{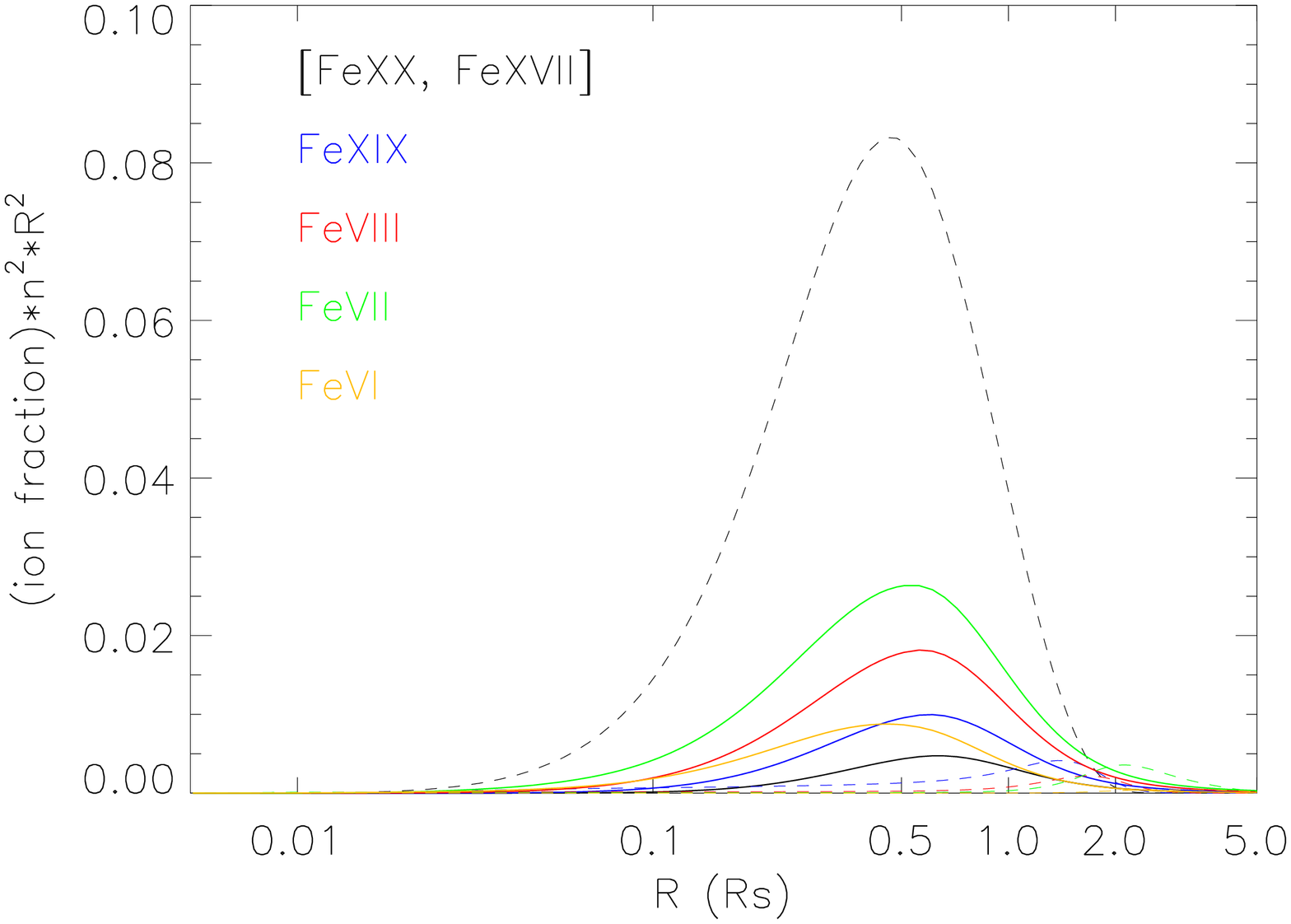}
   }
   \mbox{
  \includegraphics[width=0.5\textwidth]{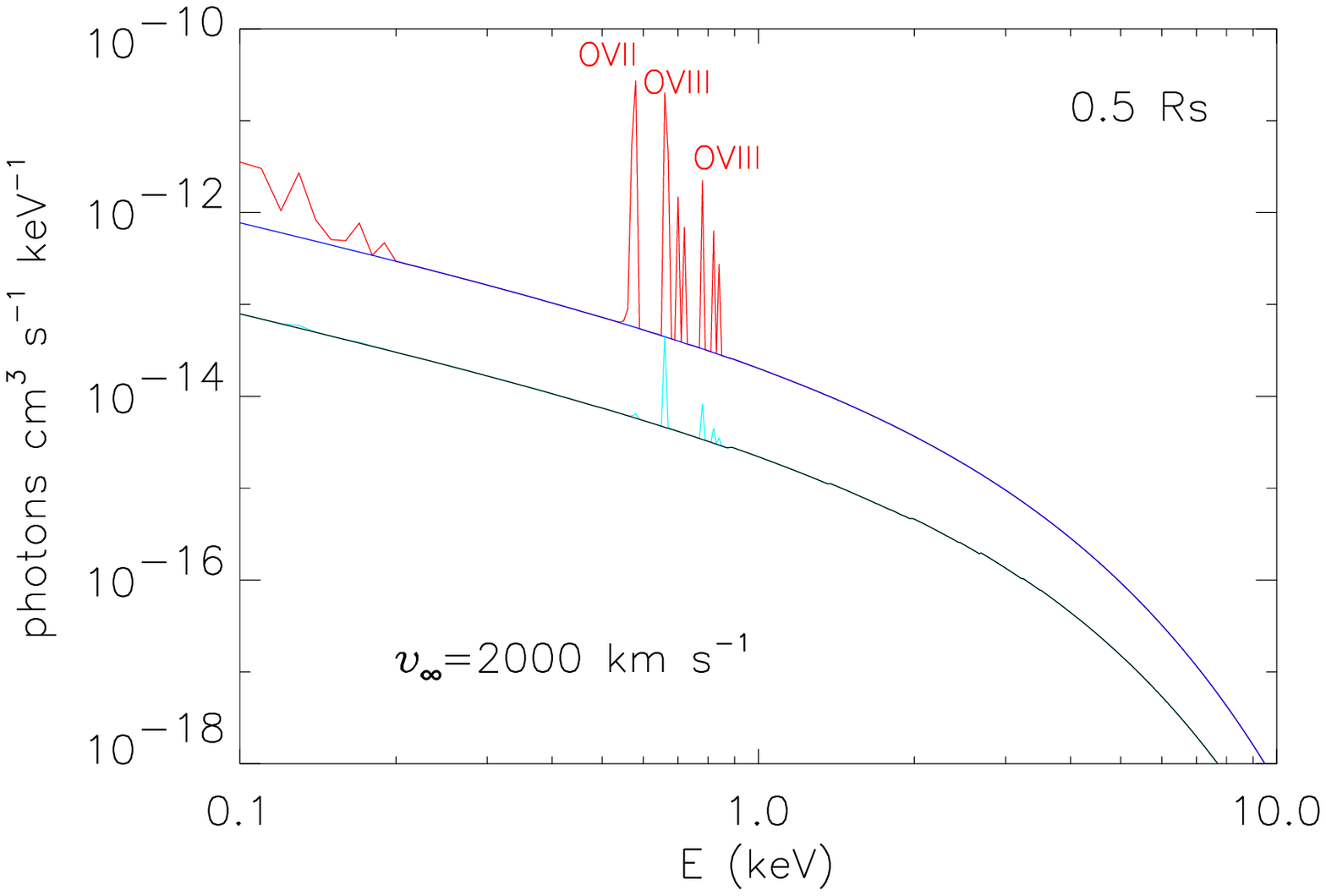}
  \includegraphics[width=0.5\textwidth]{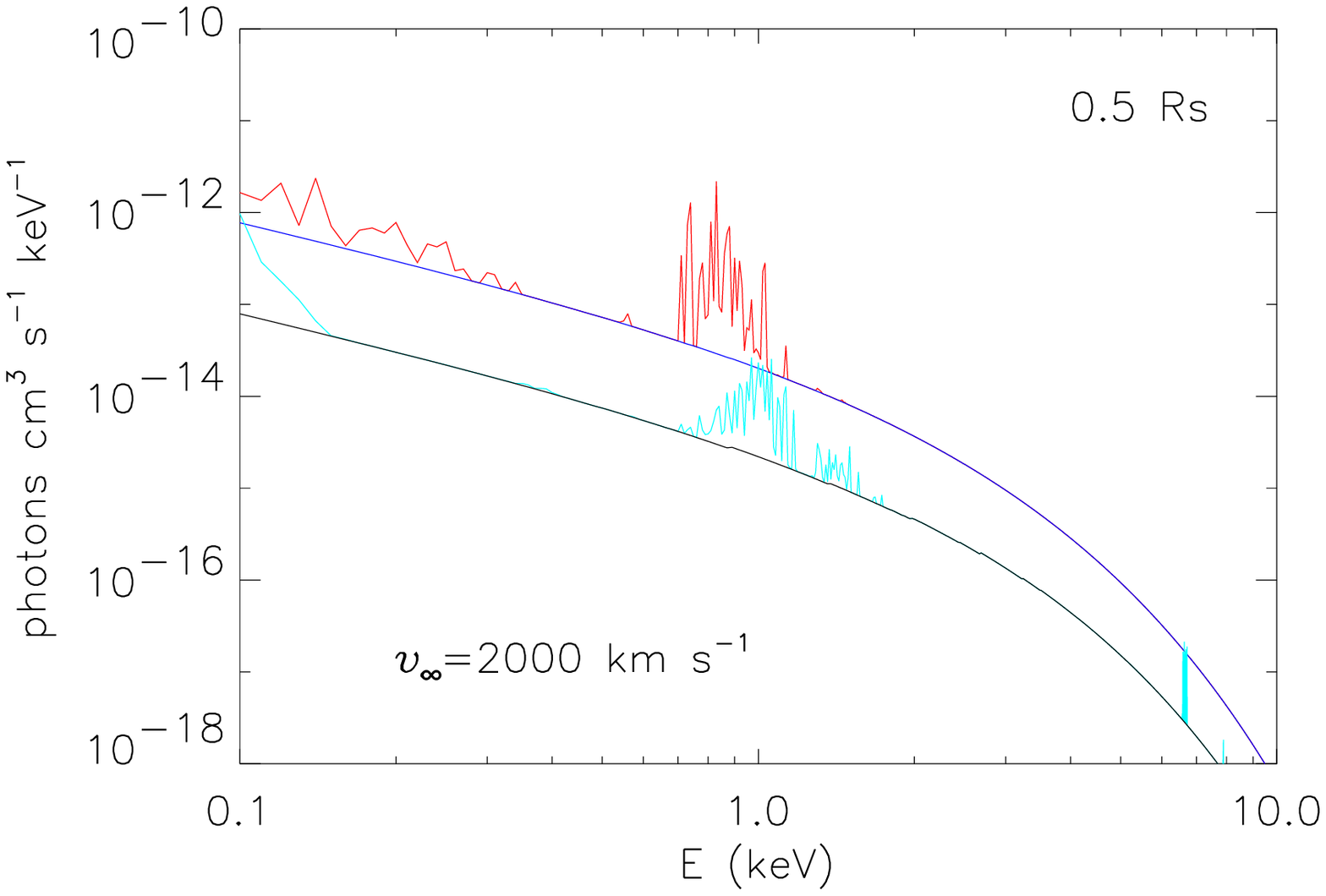}
  }
   \caption{Upper panel: Overall spectral contributions as a function of 
                 radius for Oxygen ions (left) and Fe ions (right) in the
                 CIE (dash lines) and non-CIE (solid lines) scenarios.  
                 Lower panel: Comparison between the spectral
                    contributions from Oxygen (left) and Fe (right) plus the
                   total continuum (black) at $R = 0.5 R_s$: CIE (black + cyan) and non-CIE 
                 (blue+red). For clarity, the non-CIE spectra have been multipled by 
		 a factor of 10. All are based on the wind profile as in Fig. \ref{fig-grid-m4-v2}. }
   \label{fig-neion_ionf_cumu_spec}
\end{figure}

Second, we find that the effect of the NEI manifests itself in both the inner
and outer regions
of the cluster wind, but under different  dynamical conditions.  In the
inner region it is expected
that the CIE and non-CIE calculations will be similar for the cases of high 
central densities.  Fig. \ref{fig-grid-m4-v5} illustrates such a case
($\dot{M}_{0}=1\times 10^{-4} ~\rm M_{\odot}~yr^{-1} $ and $V_{\infty}=500 ~\rm
km~s^{-1}$), in which the CIE approximation is good. 
Both calculations produce very similar cumulative spectra in either the [0,
$R_{s}$] or [$R_{s}$, 2$R_{s}$] region, although the 
count intensity distributions can differ by as much as a factor of  
2 in either the soft or hard X-ray band.
On the other hand,  the assumption of CIE in the inner region is poor for 
$\dot{M}_{0}=1\times 10^{-4} ~\rm M_{\odot}~yr^{-1} $ 
and $V_{\infty}=2000 ~\rm km~s^{-1}$, 
which leads to a low central density.
Fig. \ref{fig-grid-m4-v2} shows clearly that the two ionization calculations
produce quite different spectra. Also the soft X-ray count intensity from 
the CIE calculation is  an 
order of magnitude weaker than that from the non-CIE calculation. This discrepancy 
is due to the difference in the ionic fraction of the two calculations (Fig.
\ref {fig-neion_ionf_com}). The ionization time scale is longer  than the dynamical 
time scale over which the density drops, so O and Fe are not ionized to as high
a degree as specified by CIE at the high central temperature. 
We refer to this contrast as delayed
ionization.  Fig. \ref{fig-neion_ionf_cumu_spec}
shows  the spatial dependence of the ionic fraction weighted by $n^{2}R^{2}$ 
pertinent to the calculation of the cumulative spectrum, and 
compares the CIE and 
non-CIE spectra at $R=0.5~  R_{s}$, where the largest contribution to the
cumulative spectrum for 
[0, $R_{s}$] comes from. Clearly,  much stronger OVII, OVIII lines and Fe-L
complex show up in the non-CIE 
spectrum.   They contribute to the soft X-ray excess seen in Fig. \ref{fig-grid-m4-v2}.

% non-CIE first case, delayed recombination:
%figure 10
% weak case
%
%strong case
%figure 11
 
\begin{figure}[h]
    \centering
    \mbox{
  \includegraphics[height=0.5\textwidth,angle=90]{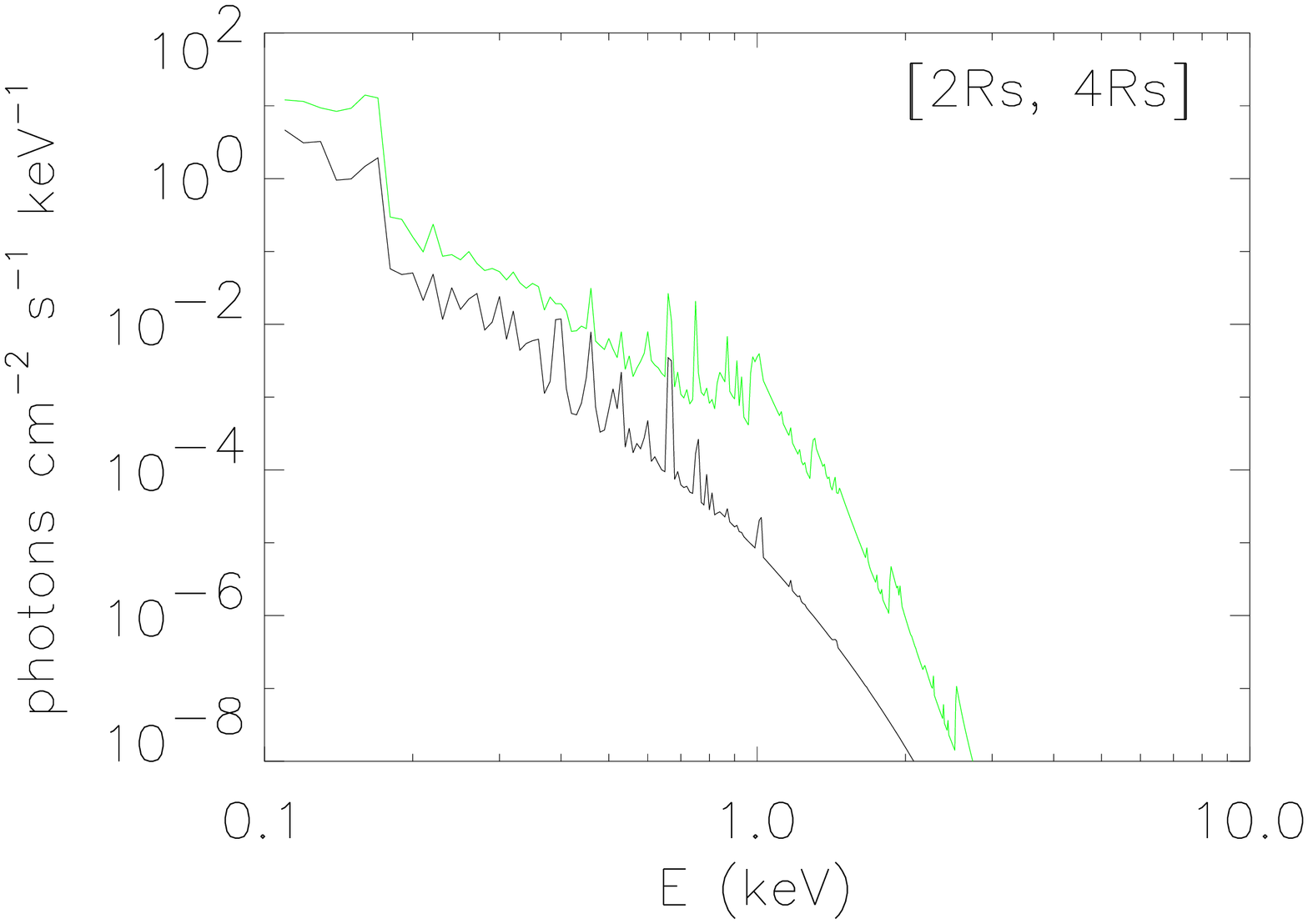}
  \includegraphics[height=0.5\textwidth,angle=90]{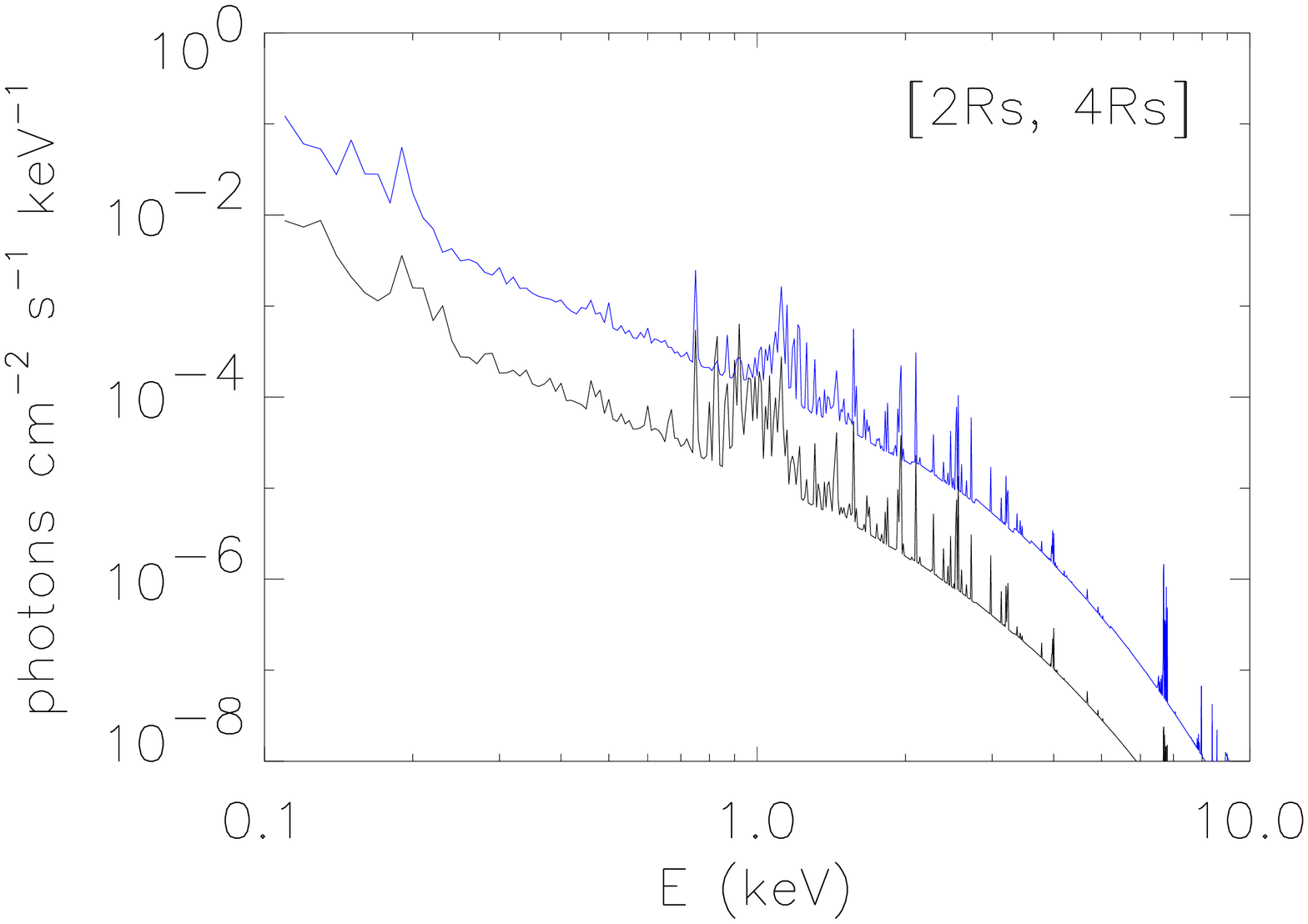}
   }
\caption{Cumulative spectra of the cluster wind with $\dot{M}_{0}=10^{-3} ~\rm M_{\odot}~yr^{-1} $ and 
		 ~ $V_{\infty}=500~\rm(left~ panel, green) ~or ~2000~\rm (right~ panel, blue)~ km~s^{-1}$.
		  For comparison,  the corresponding CIE calculations (black) are also shown. 
               For clarity, each non-CIE spectrum has been multipled by a factor of 10.
	       }
\label{fig-grid-m3-delay}
\end{figure}

\begin{figure}[h]
   \centering
   \mbox{
  \includegraphics[width=0.47\textwidth]{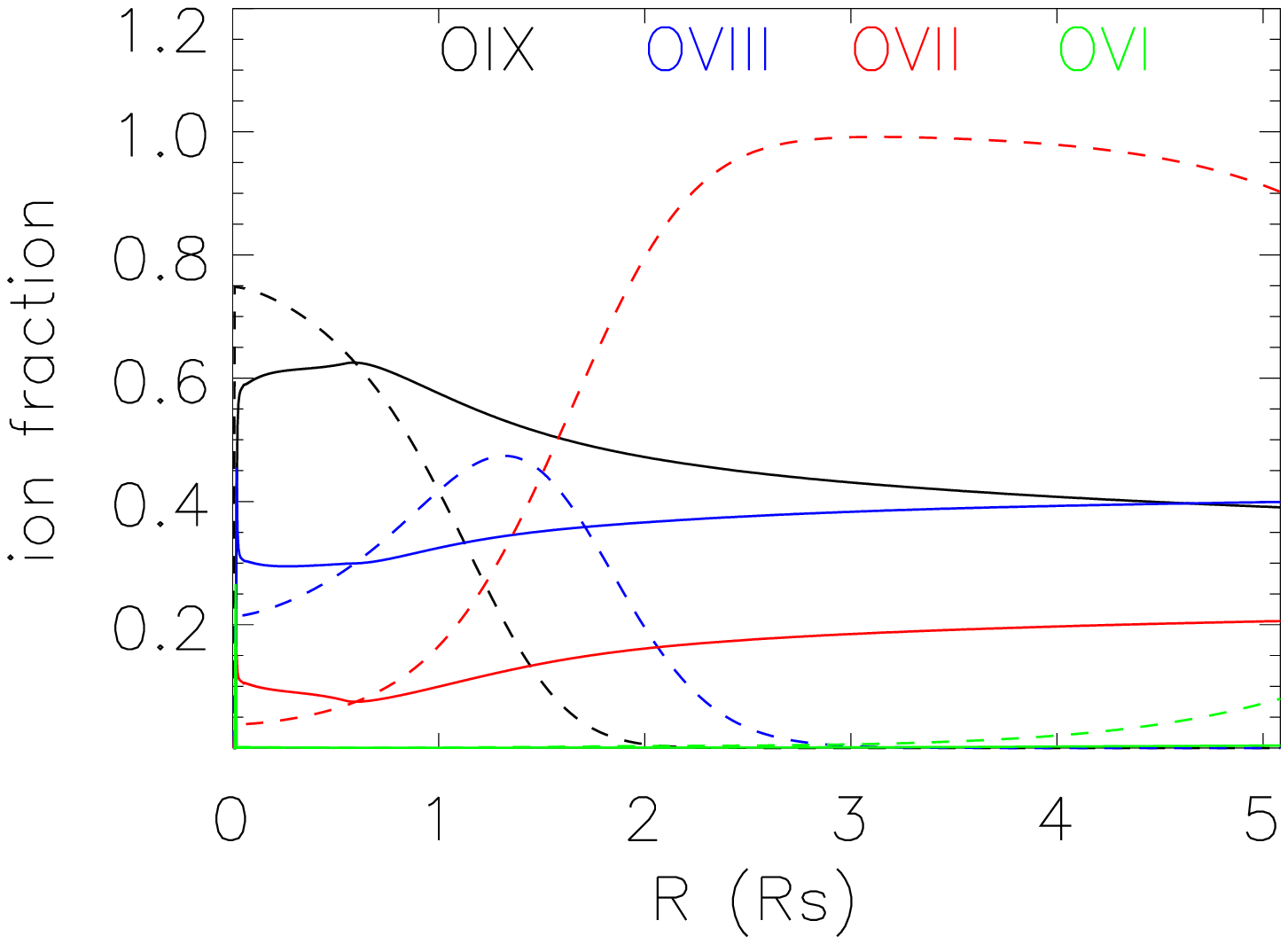}
   \includegraphics[width=0.53\textwidth]{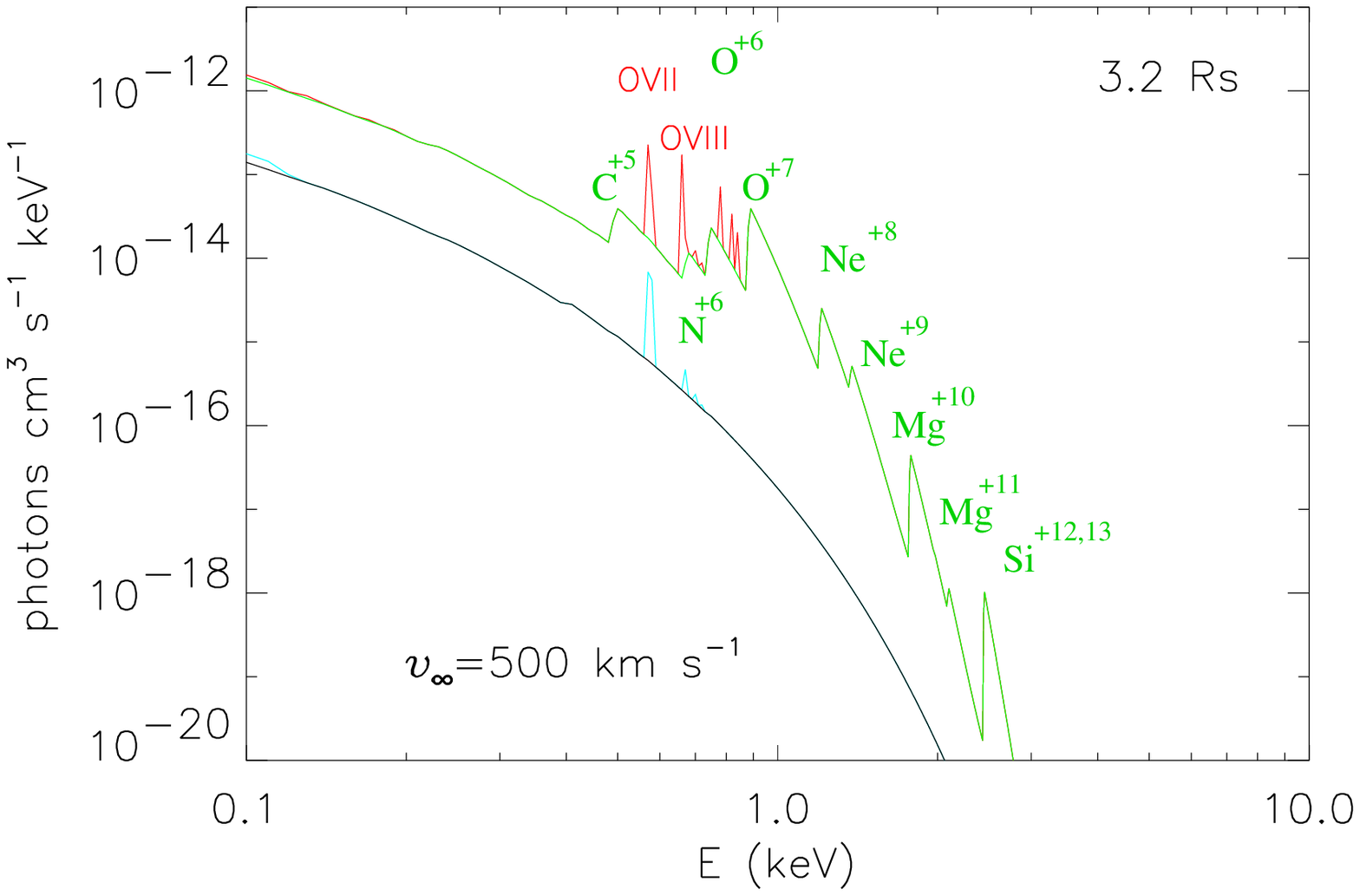}
   }
   \caption{ Left panel: Comparison of the Oxygen ionic fractions between 
                   the CIE (dash) and non-CIE (solid) calculations of the cluster wind with
                    $\dot{M}_{0}=1\times 10^{-3} ~\rm M_{\odot}~yr^{-1} $ 
                   and $V_{\infty}=500~\rm km~s^{-1}$..
                  Right panel: Similar to the lower left panel of Fig.
                      \ref{fig-neion_ionf_cumu_spec} but at 3.2 $R_{s}$.
                    Free-bound recombination edges are denoted by the contributing ions (green).
                 }  
  \label{fig-neion_ionf_com_spec}
\end{figure}

In the outer region, specifically [2$R_{s}$, 4$R_{s}$], the density has
decreased so much from spherical expansion that 
the CIE assumption is clearly not valid in any of the six wind profiles
studied. The contrast between the 
non-CIE and CIE spectra then depends on how far the temperature 
drops below $\sim 3\times 10^{6} ~\rm K$,
when OIX ions begin to  recombine.  The more rapidly the temperature
decreases with radius, the more the ionization structure deviates 
 from the CIE result. Fig. \ref{fig-grid-m3-delay} compares the CIE and non-CIE spectra in the
[2$R_{s},$ 4$R_{s}$] region. While the two spectra are quite similar
in the $V_{\infty}=2000 ~\rm km s^{-1}$ case, the  non-CIE spectrum in the $V_{\infty}=500 ~\rm km s^{-1}$
case clearly produces a
stronger hard X-ray flux from a saw-tooth-like recombination 
continuum.  The left panel of Fig. \ref{fig-neion_ionf_com_spec} compares
Oxygen ionic fractions between CIE and non-CIE calculations.  In the outer region
(e.g., $r= 3.2 R_{s}$),
OIX, which produces $\rm O^{+7}$ recombination edges and cascade lines,
is absent in the CIE but present in the non-CIE calculation. 
The right panel of Fig. \ref{fig-neion_ionf_com_spec} clearly shows 
the corresponding spectral differences in the two calculations.
%
% explanation for the soft excess
%figure 12

%figure 13
%
%explanation
%figure14

%figure 15
\begin{figure}[h]
   \centering
   \includegraphics[width=0.8\textwidth]{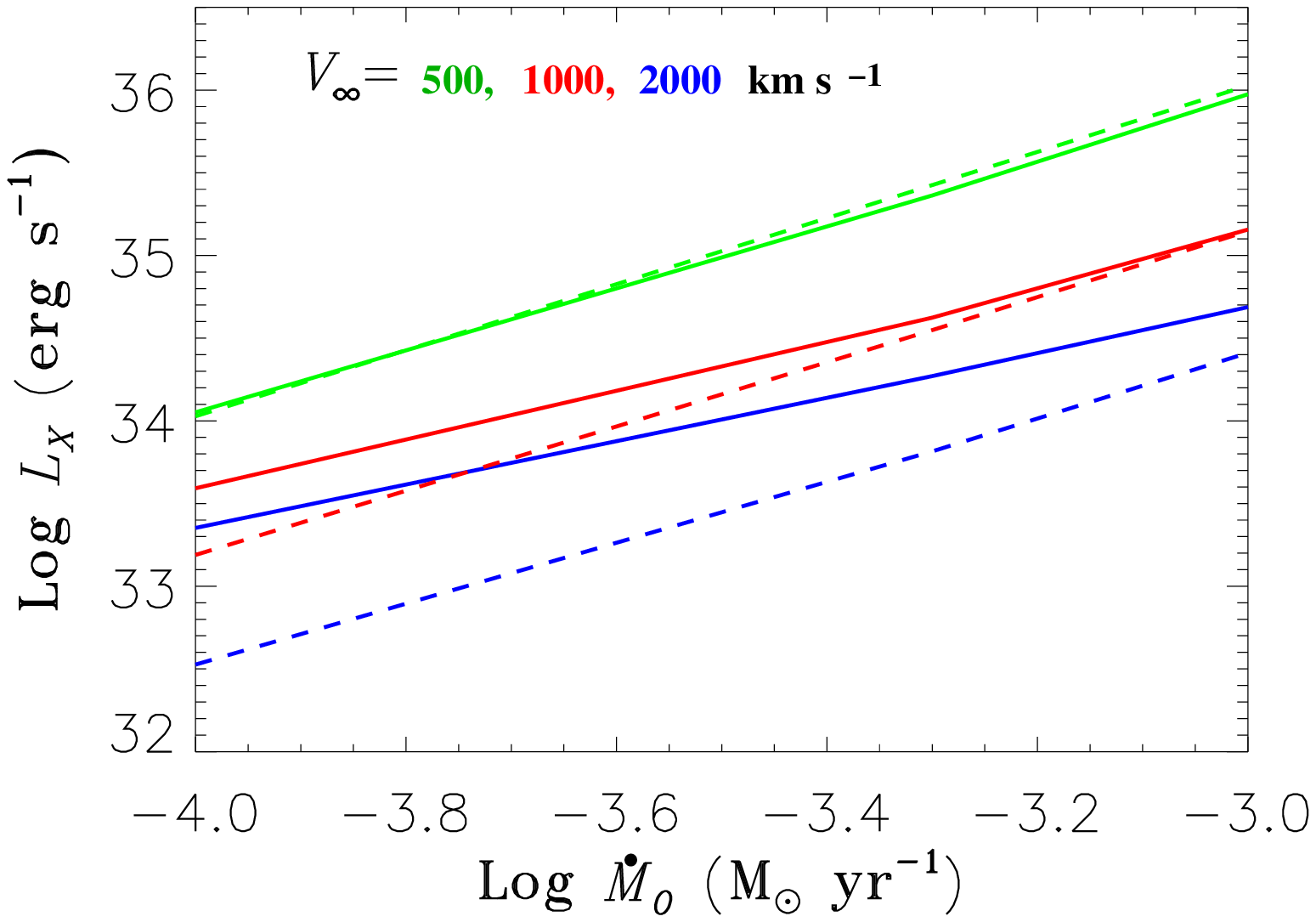}
   \caption{
        CIE (dash) and non-CIE (solid) calculations of the cumulative X-ray luminosity 
         versus mass input rate. An exponential stellar mass distribution with $R_{s}=1.97$ pc
        and $V_{\infty}=500\rm~ (green),~1000 \rm~(red), ~and ~2000 \rm~(blue) ~km ~s^{-1}$
	is studied.
        }
   \label{fig-neion_lum_m_ssc}
\end{figure}

Fig. \ref{fig-neion_lum_m_ssc} shows the dependence of the cumulative X-ray
luminosity on the mass input rate and wind velocity. The large discrepancy 
(up to almost one magnitude) between the CIE and non-CIE calculations happens for the 
high $V_{\infty}$ and low  $\dot{M}_{0}$ combination, i.e., at a high ratio of energy
to mass input rate.  
In addition, the slope dependence is exactly 2 in the CIE calculation in which 
the cooling function $\Lambda$ depends only 
on the local temperature and density. At 
a fixed $V_{\infty}$, the temperature distribution is insensitive to 
$\dot{M}_{0}$, so $L \propto \Lambda n^{2} \propto \dot{M}_{0}^{2}$. 
In the non-CIE case, however, $\Lambda$ depends on the evolutionary history of
the ionization structure; the slope varies from 2 for
$V_{\infty}=500 ~\rm km~s^{-1}$, in which the CIE approximation
is valid,  to $\sim 1.3$ for $V_{\infty}=2000 ~\rm km~s^{-1}$. 

The above analysis indicates that the ionization structure is intimately related
to the dynamical evolution. CIE is a fair 
approximation only for the case of a low ratio of energy to mass input rate.
Some other processes not considered in our model might weaken/strengthen the 
non-CIE effect for the stellar cluster wind scenario.
For example, outflows from protostars present in a cluster can be regarded 
as a mass loading process.  
In this case, the total mass input rate and the density will increase. As shown
in SH03, the velocity and temperature will decrease instead, which causes a lower
ratio of energy to mass input rate.

Our model may also be applied to galactic superwinds. X-ray emission from strong
superwinds may be detectable  at very larger radii, where delayed recombination
becomes important. But in this case, radiative cooling may not be neglected. 
Theoretical recombination spectrum has been shown in modelling galactic 
wind/outflow for the local starburst galaxy NGC3079 \citep{brei03}. 

\clearpage
\section{APPLICATION TO NGC 3603}
\label{sec-application}
NGC 3603 at a distance of 7 kpc is the most luminous giant HII region 
in the Galaxy \citep[e.g.,][]{moff94}. 
The star cluster responsible for the HII region is 
regarded as a Galactic analogue to SSCs observed in other galaxies.  
With an age of less than $3 \rm~Myr$ \citep{sube04}, the cluster contains
massive stars as well as
low-mass pre-main sequence (PMS) stars. 

Stars in NGC 3603 have been investigated from near-infrared 
to X-ray \citep[e.g.,][]{sube04, moff02}. Based on 44 spectroscopically
classified stars, which dominate the mass and energy injections, 
SH03 estimate the total mass input rate from the cluster
as $\dot{M}_{0}=2.3\times 10^{-4} ~\rm M_{\odot} ~\rm yr^{-1}$ 
and the mean weighted stellar wind terminal velocity as $V_{\infty}=2844 ~\rm km ~\rm s^{-1}$. 
We use the observed diffuse X-ray emission of NGC 3603 to test our 
NEI model of stellar cluster winds.

\subsection{Data Analysis}
\label{sec-obser}
We use the {\sl Chandra} ACIS-I observation of NGC 3603 (observation ID 633),
which was taken 
in May, 2000 for an exposure of 50 ks. \citet{moff02} used this observation
to study primarily discrete X-ray sources, although they also characterized 
the remaining ``diffuse''
emission with a one-temperature CIE thermal plasma model.
Here, we concentrate on the diffuse emission.

We reprocess the ACIS-I event data,
using the {\sl Chandra} Interactive Analysis of Observations software
package (CIAO, version 3.2). This reprocess, including the exclusion of
time intervals with background flares, results in a net 47.4 ks exposure 
(live time) for subsequent analysis.   

\begin{figure}[h]    
      \centerline{
      \includegraphics[width=1.0\textwidth]{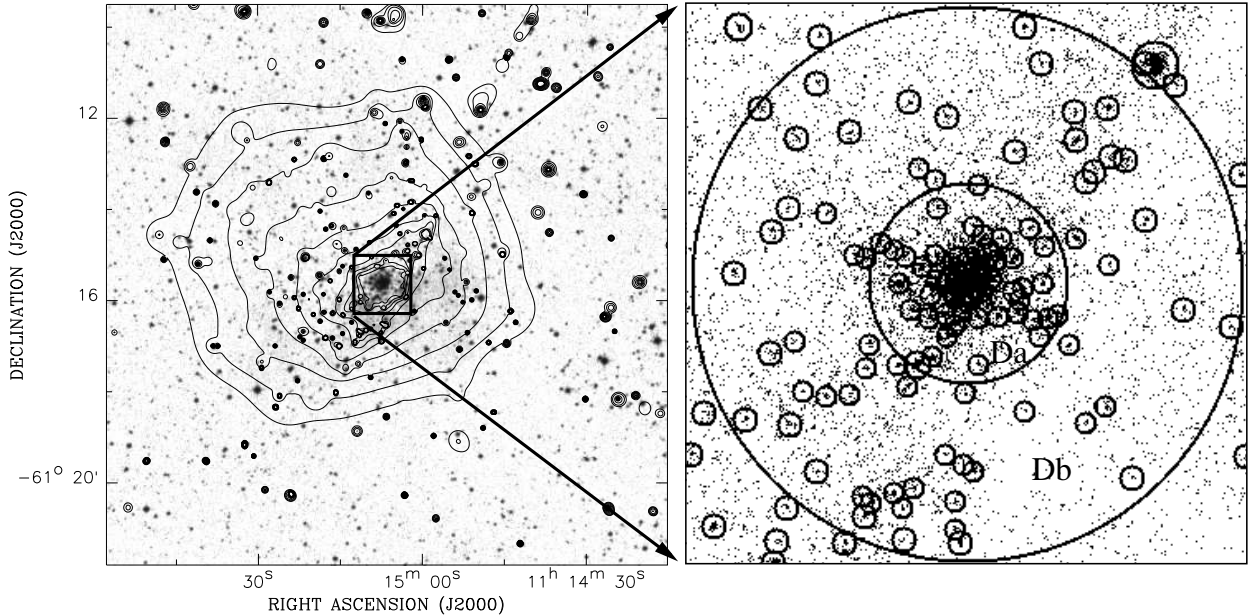}
       }
      \caption{{\sl Chandra} ACIS-I observation of NGC 3603. Left panel:~
           The 0.5-8.0 keV intensity contours (at 52, 58, 70, 94, 130, 178, 238, 310, 394, 
             and 490 $\times 10^{-4} \rm ~counts~ s^{-1}~ arcmin^{-2}$) overlaid on an optical image.
          Right panel:~ A close-up of the central cluster. The image shows the
          ACIS-I event distribution. Point-like X-ray sources 
          are marked individually with small circles of 2$\times$
         the PSF 90\% EER. The two annulii, Da ([$19\arcsec, ~30\arcsec$])
	     and Db ([$30\arcsec,~83\arcsec$]), are used to extract the spectra of 
	      the diffuse emission.  
	      }
      \label{fig-img}
   \end{figure}

The superb spatial resolution of the data allows for the detection and 
removal of much of the X-ray contribution from individual stars. We search for
X-ray sources in the 0.5-2, 2-8, and 0.5-8 keV bands 
by using a combination of three source detection and analysis
algorithms: wavelet, 
sliding-box, and maximum likelihood centroid fitting \citep{wang04}.
Source count rates are estimated with the 90\% energy-encircled radius (EER) 
of the point spread function (PSF).  The accepted sources all have the local
false detection probability $P \leq 10^{-6}$.  
The right panel of Fig. \ref{fig-img} shows the detected point sources
within the cluster field. The source detection in the inner $\sim
20^{\prime\prime}$  radius
of the cluster is highly incomplete due to severe source confusion. 

%NGC3603
%figure 16

To analyze the remaining ``diffuse" emission from the cluster, we remove the 
detected source regions from the data. 
For each faint source with a count rate (CR) $\leq 0.01\rm~ counts~ s^{-1}$, 
we exclude a region of twice the 90\%  EER, which increases with off-axis angle.  
For each source with CR $> 0.01~\rm counts~ s^{-1}$, an additional factor of
$1+\rm log (CR/0.01)$
is multiplied to the source-removal radius to further minimize the residual
contribution from the PSF wings of relatively bright sources.  

%figure 17
\begin{figure}[h]
  \centering
    \includegraphics[height=0.6\textwidth]{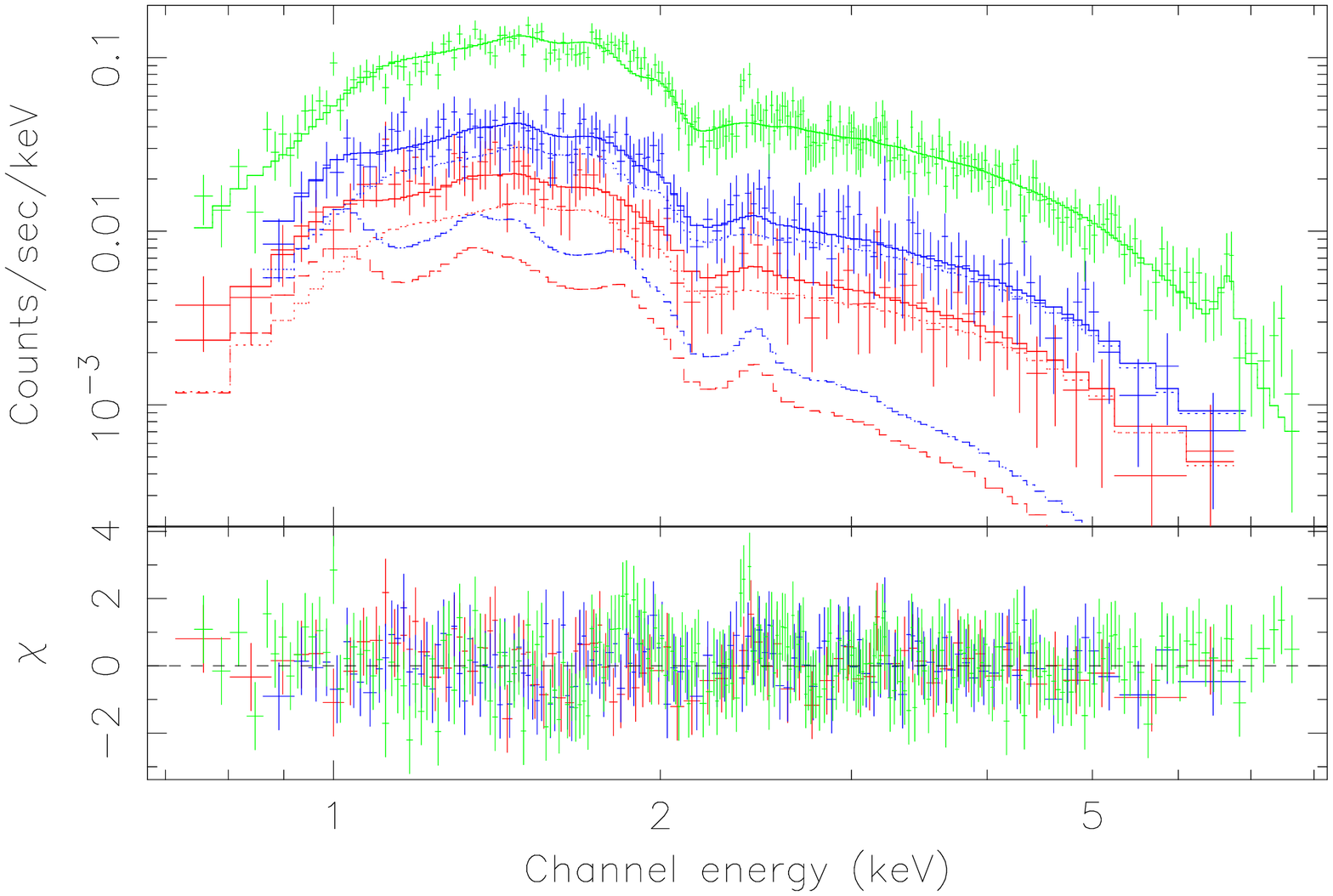} 
  \caption{ 
        Joint fit to the spectra of the point-like sources (green)
        and the two diffuse X-ray spectra (red and blue) of NGC 3603.
        All these spectra are grouped to contain at least 25 net counts per bin. 
         The point-source spectrum is modelled with an optically-thin thermal plasma, 
        while the diffuse X-ray spectral model also contains a joint-fitted residual
      point-like source contribution (dotted histograms) as well as the
       cluster wind component (dashed).  
	}
   \label{fig-fit}
   \end{figure}

\begin{figure}[h]
      \centerline{
      \includegraphics[width=0.47\textwidth]{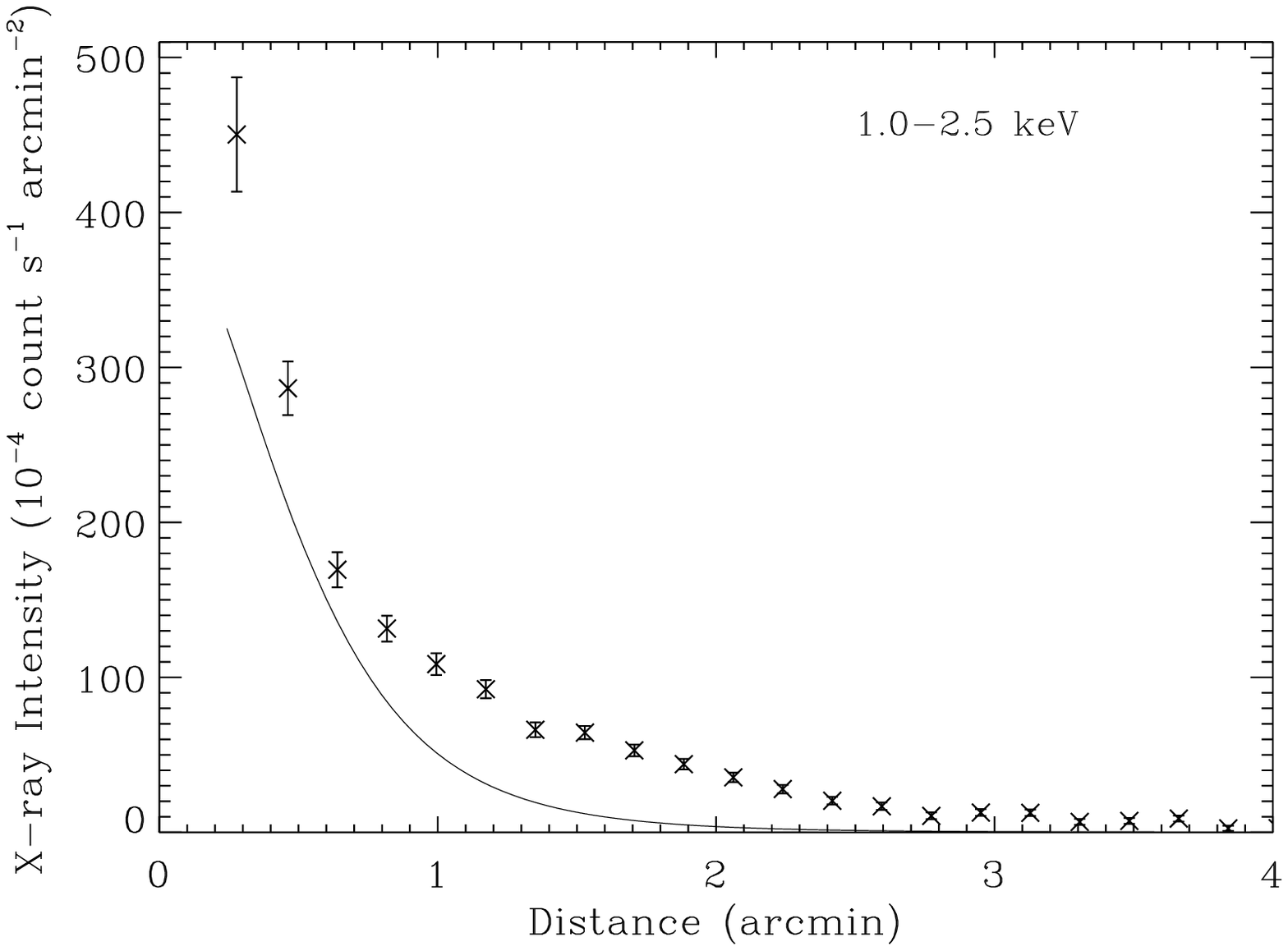}
      \hspace{0.3in}
      \includegraphics[width=0.45\textwidth]{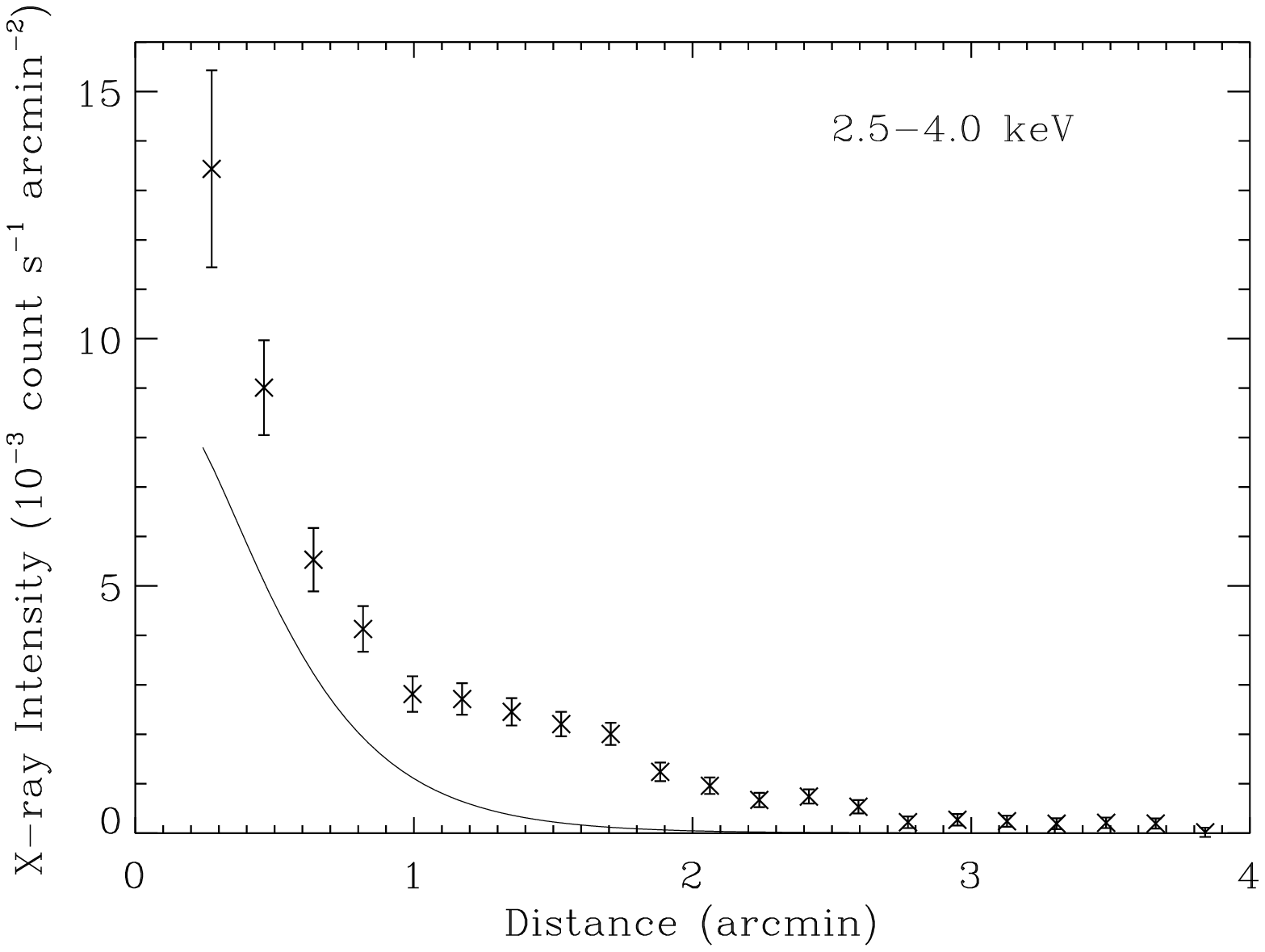}
       \vspace{-0.05in}
       }
      \caption{ACIS-I intensity profiles 
            of the diffuse X-ray emission (crosses with $1\sigma$ error bars) in the 
             1.0-2.5 keV (left) and 2.5-4.0 keV (right) bands. The solid lines
        represent the predicted cluster wind contributions based on the spectral fit.  
	}
      \label{fig-radpf}
   \end{figure}

We extract the spectra of the diffuse X-ray emission separately from the two
annular regions, Da and Db
(divided to contain roughly the same number of counts), 
as well as an accumulated point source spectrum from the combined region (Da+Db).
An additional spectrum extracted from the outer annulus between  4\farcm0 and
 5\farcm0 radii is used as the local background of the diffuse X-ray emission, 
which in turn are normalized
for the background subtraction of the sources.
Fig. \ref{fig-fit} shows a joint-fit of the spectra with our cluster wind model
plus a thermal model,
which is used to characterize the point-like source spectrum. Here we assume that 
the detected and undetected point sources have the same spectral shape.  
We jointly fit the diffuse spectra and the accumulative spectrum of point sources
together. 
We find that the spectral fit results are insensitive to the assumed stellar
distributions of our model
and to the exact scale radius. Therefore,  Table \ref{tab-fit-exp} includes only the results
from the model with the more physical exponential stellar mass distribution; the
fixed scale radius 
$r_{sc}=0.68~\rm pc$ corresponds to the same sonic radius as adopted by SH03. 
The best-fit model predicts the luminosity contributions 
of $L_{\rm 2-8~keV}\sim 2.5\times 10^{33}\rm~ erg~s^{-1}$ and $8.8  \times
10^{33}\rm ~erg~s^{-1}$
from the cluster wind and residual point-like source components.
Fig. \ref{fig-radpf} shows the predicted radial surface intensity distribution of the
cluster wind contributions compared with the observed
``diffuse" emission in two energy bands.  

%figure 18

%%%%%%%%%%%%%%%%%%%%%%%%%%%%%%%%%%%%%%%%%
%table for exponential case
%%%%%%%%%%%%%%%%%%%%%%%%%%%%%%%%%%%%%%%%%
\begin{deluxetable}{cc|c}
\footnotesize
\tablecolumns{3}
\tablecaption{Joint spectral fit results\tablenotemark{a} \label{tab-fit-exp}}
\tablewidth{0pt}
\tablehead{
\multicolumn{2}{c}{Parameter} & Model
}
\startdata
${N_{\ssst\rm H}}$& $ (10^{22}~\rm cm^{-2})$& $0.81^{+0.06}_{-0.05} $    \\  \hline
\multicolumn{3}{c}{Cluster wind component}    \\ 
%\multispan{Cluster wind component}\\
$r_{sc}$  &  (pc)  & 0.68 (fixed) \\
Abundance & $(Z_{\odot})$  & $ 0.21^{+0.12}_{-0.11}$     \\
$Log \dot{M}_{0}$ &$(\rm M_{\odot}~\rm yr)$ &$-3.37^{+0.42}_{-0.24} $  \\
$V_{\infty}$  & ($ 1\times 10^{3} \rm ~km ~s^{-1}$) &1.53 ($> 0.50$)\\ \hline
\multicolumn{3}{c}{Point-like source component}   \\ 
%\multispan{Point-like source component}\\
$kT$     & (keV) &$3.79^{+0.41}_{-0.33}$ \\
$K_{1} $\tablenotemark{b} & ($10^{-4}$ photons cm$^{-2}$)  
&$3.63^{+0.77}_{-1.56} $  \\ 
$K_{2} $\tablenotemark{c} & ($10^{-4}$ photons cm$^{-2}$) &
$7.76^{+1.38}_{-1.73} $ \\ 
$K_{3} $\tablenotemark{d} & ($10^{-3}$ photons cm$^{-2}$)  &
$3.26^{+0.22}_{-0.19} $ \\ \hline
$\chi^{2}/{\rm d.o.f.}$ & &$301.3/457 $  \\ 
\enddata
\tablenotetext{a}{Note: all error bars are at the 90\% confidence}
\tablenotetext{b}{$K_{1} $ is the MEKAL  normalization of the undetected point
sources in Region Da.}
\tablenotetext{c}{$K_{2} $ is same as $K_{1} $, but in Region Db.}
\tablenotetext{d}{$K_{3} $ refers to the MEKAL normalization of the detected point
sources.}
\end{deluxetable}

\subsection{Discussion}
\label{sec-discuss}
The above results represent the first quantitative comparison of a self-consistent 
NEI cluster wind model with observational data. 
Our fitted foreground X-ray-absorbing gas column density,
$N_{H}=8.1^{+0.6}_{-0.5} \times 10^{21}\rm ~cm^{-2}$, is  consistent 
with the interstellar reddening toward NGC 3603.
The reddening of $E(B-V)= 1.25 - 1.8 \rm~ mag $, determined
by \citet{sube04} from 115 bright stars in the field, corresponds to 
$N_{H}=(7.3 - 10.4)~\times 10^{21}\rm ~cm^{-2}$ \citep[e.g.,][]{bohl78}.
The derived abundance of the wind is sub-solar: ${\rm Z}=0.21^{+0.12}_{-0.11}~Z_{\odot}$.
This abundance likely represents an underestimate of the true value due to the 
oversimplification in the 1-D model, 
which does not account for substructures in both temperature and density. 
Such substructures lead to smearing out of spectral line features that  
characterize the metal abundance. Our fitted mass input rate and terminal velocity
of the cluster wind is, within a factor of $\sim 2$,  consistent with the empirical 
estimation by SH03 (\S \ref{sec-application}).  
But the cluster wind contributes only about 25\% of the observed total diffuse
emission luminosity. 

While the remaining of the diffuse emission likely originate from undetected
point-like sources, what is their nature?  \citet{moff02} and \citet{sube04} 
suggested that PMS young stellar objects (YSOs) may be responsible.  
According to the study by \citet{saga01}, the NGC 3603 cluster has an 
initial mass function (IMF) that can be 
characterized by a power law with a slope of 0.84 and contains about $2\times
10^{2}$ stars in the mass range 7-75 $\rm M_{\odot}$. 
Using this IMF, we estimate the presence of about  
$8\times 10^{3}$ YSOs (0.3-3 $\rm M_{\odot}$) in this cluster. 
The study of such YSOs in the Orion Nebular \citep{feig05} shows that their average
2-8 keV luminosity is $\sim 1 \times 10^{30} \rm erg~s^{-1}$ per star. 
Thus the expected total YSO contribution
in the NGC 3603 cluster is $L_{\rm 2-8~keV}\sim 8  \times 10^{33}$, 
consistent with the prediction from our spectral fit.  

\section{SUMMARY}
In this paper, we have presented a NEI spectral code, which is based on our
updated atomic data, 
including the atomic process of recombinations into highly excited levels. We
have shown important differences between the CIE and non-CIE calculations.

We have constructed a self-consistent physical model for super stellar cluster
winds, combining the NEI code with a 1-D steady-state adiabatic wind
solution. We find that in the inner region of the star cluster, the NEI 
effect is significant in the regime of a high energy to mass input
ratio and manifests in an enhanced soft X-ray emission, which could be an order
of magnitude more luminous than that obtained under the CIE assumption.

We apply the cluster wind model to the {\sl Chandra} data on NGC 3603 
and find that the inferred mass and energy input rates of the stellar cluster
winds are comparable 
to the empirical estimates. But the cluster wind in NGC 3603 accounts for no 
larger than 50\% of the total diffuse X-ray emission. YSOs are likely
responsible for the remaining diffuse emission. 

\acknowledgements  We are grateful to Z. Li for providing the software programs
used for calculating the wind dynamics.
We thank K. J. Bowkowski, M.-F. Gu,  J. S. Kaastra, F. Paerels, and R. Smith for
many valuable comments and suggestions on this work, which 
is supported by NASA through the grant SAO/CXC GO4-5010X.

\end{document}